\documentclass{aa}  
\usepackage{amstext}
\usepackage{amsmath,bbold,mathrsfs,bm,mathtools}
\usepackage{graphicx}
\usepackage{textcomp}
\usepackage[varg]{txfonts}
\usepackage{time}
\usepackage{subfig}

\usepackage{url}
\usepackage{upgreek}

\usepackage{natbib}
\bibpunct{(}{)}{;}{a}{}{,}

\usepackage{url}
\usepackage{rotating}
\usepackage{float}

\usepackage{multirow} 
\usepackage{array} 
\usepackage{balance}
\usepackage[table]{xcolor}
\usepackage[breaklinks=true]{hyperref} %% to avoid \citeads line fills

\usepackage{xifthen}

\makeatletter
\def\url@leostyle{%
  \@ifundefined{selectfont}{\def\UrlFont{\sf}}{\def\UrlFont{\tiny\ttfamily}}}
\makeatother
\begin{document}

\title{Off-disk straylight measurements for the\\ Swedish 1-meter
  Solar Telescope}

% \title{Very wide-wing straylight measurements for\\ the Swedish 1-meter
% Solar Telescope}

\author{Mats G. L{\"o}fdahl}  

\institute{Institute for Solar Physics, Dept. of Astronomy, Stockholm
  University, Albanova University Center, 106\,91 Stockholm, Sweden}

% \date{Received September 15, 1996; Accepted March 16, 1997}
\date{Draft: \now\ \today} 

\frenchspacing 

\abstract{Accurate photometry with ground-based solar telescopes
  requires characterization of straylight. Scattering in
  Earth's atmosphere and in the telescope optics are potentially
  significant sources of straylight, for which the point spread
  function (PSF) has wings that reach very far. This kind of
  straylight produces an aureola, extending several solar radii off
  the solar disk.}
{We want to measure such straylight using the ordinary science
  instrumentation.}
{We scanned the intensity on and far off the solar disk by use of the
  science cameras in several different wavelength bands on a day with
  low-dust conditions. We characterized the far wing straylight by
  fitting a model to the recorded intensities involving a
  multicomponent straylight PSF and the limb darkening of the disk.}
{The measured scattered light adds an approximately constant fraction
  of the local granulation intensity to science images at any position
  on the disk. The fraction varied over the day but never exceeded a
  few percent. The PSFs have weak tails that extend to several solar
  radii, but most of the scattered light originates within
  $\sim$1\arcmin.} 
{Far-wing scattered light contributes only a small amount of
  straylight in SST data. Other sources of straylight are primarily
  responsible for the reduced contrast in SST images.}
% 5 {} token are mandatory

%%% \abstract
%%%% context heading (optional)
%%%% {} leave it empty if necessary  
%%% {}
%%%% aims heading (mandatory)
%%% {}
%%%% methods heading (mandatory)
%%% {}
%%%% results heading (mandatory)
%%% {}
%%%% conclusions heading (optional), leave it empty if necessary
%%% {}

\keywords{instrumentation: miscellaneous -- methods: observational --
  methods: data analysis -- techniques: photometric}

\maketitle

\section{Introduction}
\label{sec:introduction}

Careful measurements of the point spread function (PSF) in the Solar
Optical Telescope (SOT) on the Hinode spacecraft
\citep{2008a&a...487..399w} and comparison with synthetic data from
magnetohydrodynamic (MHD) simulations \citep{2009a&a...503..225w} have
established that the granulation contrast in the synthetic data is
correct. However, the contrasts measured in data from ground-based
solar telescopes is far below that of the MHD data.

Like all of today's major ground-based solar telescopes, the Swedish
1-meter Solar Telescope \citep[SST; ][]{scharmer03new}, located in the
Roque de Los Muchachos Observatory (ORM) on La Palma, has an adaptive
optics \citep[AO; ][]{scharmer03adaptive} system that corrects for
image degradation from wavefront aberrations caused by atmospheric
turbulence. Science images are also routinely restored for residual
aberrations by use of multi-frame blind deconvolution \citep[MFBD;
][]{lofdahl02multi-frame} techniques like phase diversity
\citep{lofdahl94wavefront} and multi-object MFBD
\citep[MOMFBD;][]{noort05solar}. Both AO and MFBD techniques correct
the wavefront aberrations only partially, a consequence of their being
based on a finite number of AO mirror electrodes and on an expansion
of the wavefront aberrations in a finite number of modes, leaving a
tail of uncorrected high-order modes. If the correction is of
sufficiently high order, the result is a diffraction-limited
resolution but a Strehl ratio that is reduced due to a PSF ``halo'' of
speckle noise from the the uncorrected modes. This effect has already
been simulated for stars, as well as for solar images, by
\citet{1989JOSAA...6...92S}. \citet{1998aoat.book.....H} discusses the
effect, as do several authors represented in the collection edited by
\citet{roddier99adaptive}.
\citet{scharmer10high-order} find that a significant part of the
contrast reduction in SST data can be explained by the uncorrected
modes. They also implemented a method for correcting the solar image
contrast for this effect based on the known statistics of atmospheric
turbulence and simultaneous measurements of Fried's parameter $r_0$
from a wide-field wavefront sensor. 

\citet{scharmer11detection} demonstrate that the granulation contrast
known from synthetic data can be used together with umbral intensities
to constrain both the strength and the width of the straylight PSF.
They conclude that the dominating straylight of the SST at the time of
their observations was in the form of PSFs with narrow wings
($\sim$1\arcsec{} FWHM).

Continuing the search for contrast-reducing straylight in the SST,
\citet{lofdahl12sources} measured straylight originating in the
post-focus optics of the SST. High-order modes from the deformable
mirror (DM) contributed narrow-wing straylight (kernels with 90\%
enclosed energy within $\sim$0.6\arcsec{}) with a magnitude ranging
from 34\% of the total intensity at 390~nm to 18\% at 854~nm, ghost
images about $\sim$1\%, and an unidentified diffuse component that
fitted well to a model with one or a few tenths of a percent of the
intensity in 16\arcsec\ (blue) to 34\arcsec\ (red) wide wings. 

The above results are based on data from different years and therefore
cannot easily be combined into a current straylight budget for the
SST, because several key optical components have recently been
replaced, including the tip-tilt mirror and an upgrade of the AO from
37 to 85 electrodes. We refer the reader to \citet{scharmer_sky}, who
assess the current performance of the SST based on measurements of
granulation contrast in a variety of wavelengths. We are led to a
picture where the straylight from the post-focus optics has only small
contributions from straylight with wings wider than $\sim$1\arcsec{}.
However, the measurements do not include contributions from the
telescope itself or scattering from aerosols suspended in the
atmosphere. These sources can produce straylight with a much wider
influence. \citet{zwaan65sunspot} estimated that the contributed
straylight is 4--10\% of the intensity at disk center (DC) in
measurements of the aureola with several different solar telescopes in
the fifties and sixties. However, this was for sites and telescopes
from an earlier era, and it seems likely that modern solar telescopes
have better intrinsic straylight properties. Here, we aim to
characterize the SST far wing straylight (FWS) by scanning the radial
intensity distribution, both the limb-darkened intensity on the solar
disk and the off-disk aureola, extending far outside the limb.
 
Radial intensity scanning has a long history with several different
solar telescopes. \citet{1970SoPh...12..328S} scanned from $0.9R_\sun$
to approximately $3R_\sun$ with photometers in several wavelength
bands on an off-axis mirror system of the Oslo Solar Observatory,
finding scattering with $\sim$100\arcsec{} wide wings in the visible.
\citet{1977SoPh...51...25P} measured the solar limb darkening with a
spectrometer ``as the image drifted by diurnal motion across the
entrance aperture.'' They were also apparently the first to fit limb
darkening data to fifth-order polynomials in $\mu$ (the cosine of the
viewing angle), a type of fit that we also use.
\citet{1990SoPh..125..211M} and \citet{1992SoPh..140..207M} used a
photometer to make similar drift scans with the Vacuum Newton
Telescope at the Observatorio del Teide in Tenerife. They formulated
and solved the radiative transfer problem for Earth's atmosphere to
model its scattering effects. The latter paper includes an overview of
the 1992 state of art of measuring scattering and instrumental
straylight for solar telescopes. We have not found any more recently
published measurements of the scattered light PSF in high-resolution
ground-based solar telescopes. In contrast, the major space-borne
telescopes have recently had their scattered light characterized by
\citet{2007A&A...465..291M} (SoHO/MDI), \citet{2008a&a...487..399w}
and \citet{2009A&A...501L..19M} (Hinode/SOT/BFI),
\citet{2009ApJ...690.1264D} (TRACE), and \citet{2014A&A...561A..22Y}
(SDO/HMI).

In this paper we measure the radial intensity distribution from disk
center out to several solar radii using the available science cameras
and fit the measurements to a model involving both the straylight PSF
and the solar limb darkening, as well as several other model
parameters. These fits allow us to characterize the FWS PSF and its
contribution to science data intensity on the entire disk. We describe
the collection and radial binning of our data, and discuss the
off-disk intensity in Section~\ref{sec:observations-preprocessing}. In
Section~\ref{sec:straylight-analysis} we describe the model fitting.
We discuss the results in Section~\ref{sec:conclusion}.

\section{Measurements and binning}
\label{sec:observations-preprocessing}

\subsection{Optics and instrumentation}
\label{sec:optics}

The primary optical element of the SST is a singlet lens with a focal
length of 20.3~m at 460~nm and a 98-cm aperture. A mirror at the
primary focus reflects the light to a Schupmann corrector, which forms
an achromatic focus through a field lens (FL) next to the primary
focus.

\begin{figure}[!t]
  \centering
  \includegraphics[viewport=138 426 406 675,clip,width=\linewidth]{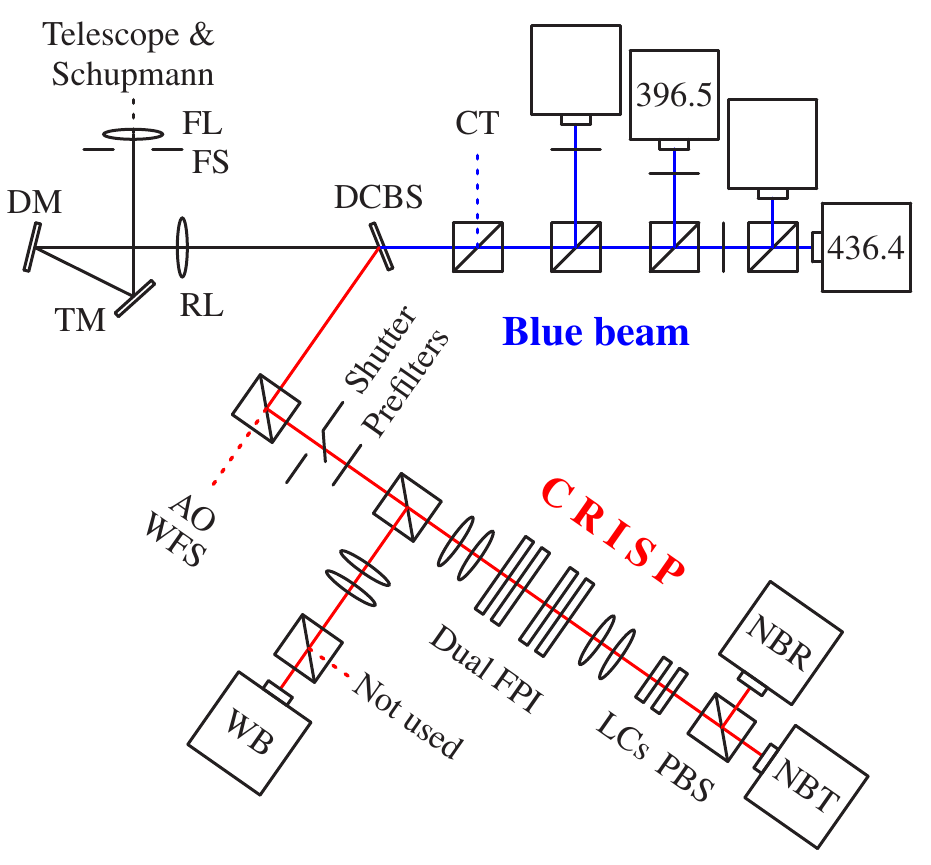}
  \caption{Setup schematics. Light from the telescope enters through
    the target from upper left. TM = tip-tilt mirror; DM = deformable
    mirror; RL = reimaging lens; DCBS = dichroic beamsplitter; CT =
    correlation tracker; WFS = wavefront sensor; LCs = liquid crystal
    analyzer; FPI = Fabry--P\'erot interferometer, PBS = polarizing
    beamsplitter, WB = wide band, NBT = narrow band transmitted, NBR =
    narrow band reflected.}
  \label{fig:setup}
\end{figure}

The setup following the Schupmann focus is illustrated in
Fig.~\ref{fig:setup}, in the caption of which acronyms for many
optical elements are defined. The FL makes a pupil image in focus on
the 85-electrode monomorph DM. The RL makes a converging F/46 beam
parallel to the optical table. This beam is split by the 500~nm DCBS
into a blue beam and a red beam. Both beams have several science
cameras behind a variety of interference filters.

We did these observations in six different wavelength bands, from
396~nm in the violet to 854~nm in the near IR. We used two cameras
with separate filters in the blue beam ($\lambda<500$~nm) and four
different CRISP \citep{scharmer08crisp} prefilters in the red beam.

The three CRISP Sarnoff CAM1M100 camera detectors are 1024 by 1024
pixels with an image scale of 0\farcs059/pixel, making a FOV of
60\arcsec, cadence $\sim$37~Hz. The two blue MegaPlus II es4020
cameras have 2048 by 2048 pixel detectors with image scale
0\farcs034/pixel, FOV 80\arcsec, cadence $\sim$10~Hz.

\subsection{Data collection}
\label{sec:data-collection}

Data were collected on 2014-04-27. This was a clear day with little
dust, as quantified by a TNG aerosol count\footnote{The concentration
  of airborne dust at ORM has been measured routinely since 2001 by
  the site-monitoring group of the Telescopio Nazionale Galileo (TNG),
  see \cite[and references therein]{2011MNRAS.416.1585L}. With a
  cadence of two hours, they count dust particles in a range of sizes
  from 0.3 to 10.0 µm with different scattering properties. The summed
  contribution of the different particle sizes is plotted in their web
  site \url{http://tngweb.tng.iac.es/weather/dust/}. This number can
  serve as a proxy for the amount of dust above the site because the
  vertical distribution of (Saharan) dust over the Atlantic is such
  that elevated concentrations are measurable at 2~km elevation
  \cite{alpert04vertical}. The range of such measurements for 2014 was
  roughly $10^{-2}$--$10^{2}$~\textmu{}g\,m$^{-3}$. Calima conditions
  are characterized by a count of $\ga10^1$~\textmu{}g\,m$^{-3}$.} of
0.3~\textmu{}g\,m$^{-3}$. We cleaned the telescope lens from dust a
few days before the observations. The observations are summarized in
Table~\ref{tab:observations} and described in detail below.

\begin{table}[!tbp]
  \centering
  % t         z          amc
  % 11.10884  32.31771   0.97557
  % 11.25095  30.47489   0.97851
  % 11.41096  28.42144   0.98151
  % 11.53218  26.89853   0.98357
  % 
  \caption{Scan data collected on 2014-04-27.}
  \label{tab:observations}    
  \begin{tabular}{@{}l@{\hspace{2mm}}lc@{\hspace{2mm}}c@{\hspace{2mm}}cc@{}}
    \hline\hline\noalign{\smallskip}
    \multirow{2}{*}{Type} & 
    \multirow{2}{*}{Time [UT]} & 
    \multicolumn{3}{c}{CW [nm]} & \multirow{2}{*}{$\theta_\text{z}$}  \\
    \cline{3-5}\noalign{\smallskip}
    && CRISP & \multicolumn{2}{c}{Blue beam}\\
    \noalign{\smallskip}\hline\noalign{\smallskip}
    \multirow{4}{*}{HA} & 11:03:05--11:09:58 & 854.2 & 396.5 & 436.4 & 32\fdg3  \\
    & 11:11:32--11:18:34 & 630.2 &       &       & 30\fdg5  \\
    & 11:21:22--11:27:56 & 557.6 &       &       & 28\fdg4  \\
    & 11:28:55--11:35:56 & 538.0 &       &       & 26\fdg9  \\
    \noalign{\smallskip}\hline\noalign{\smallskip}
    \multirow{4}{*}{DA} & \llap{(}11:49:06--12:15:51 & 854.2 & 396.5 & 436.4 & 23\fdg5--19\fdg1\rlap{)}\\
    & 12:33:16--12:47:38 & 630.2 &       &       & 17\fdg0--15\fdg8 \\
    & 12:50:43--13:01:58 & 557.6 &       &       & 15\fdg7--15\fdg3 \\
    & 13:48:16--14:01:04 &       & 396.5 & 436.4 & 17\fdg6--19\fdg2 \\
    & 14:05:04--14:18:24 & 854.2 &       &       & 19\fdg8--21\fdg9 \\
    \noalign{\smallskip}\hline
   \end{tabular}
  \tablefoot{The central wavelengths (CWs) of the CRISP prefilters
    are nominally one or a few tenths of a nm to the red of the
    wavelengths listed here, but tilted by small angles to center the
    passbands on the core wavelengths of the spectral lines for which
    they are usually used. $\theta_\text{z}$ denotes the zenith
    angle of the telescope pointing.  The
    11:49 scans are referred to below as the ``extra'' DA scans.}
\end{table}

The seeing was measured by the AO system while scanning on the disk
but not off disk. The quality of the $r_0$ measurements at the limb,
while scanning rapidly, is unknown but based on the numbers logged
over most of the disk we estimate that the seeing quality was very bad
during most of the limb passages ($r_0\approx 4$~cm) and never much
better than that ($r_0<9$~cm). Due to limitations of the SH WFS, the
seeing measurements of the SST AO lose their meaning for $r_0\la 4$~cm.
But even assuming $r_0 = 1$~cm, the FWHM of a long exposure PSF is
only a few tens of arcsec at 854~nm, representing less than a percent
of the solar radius.

We made ``drift scans'', by pointing far off the limb on the hour
angle (HA) angle axis, and then stopping the telescope tracking.
Letting the Sun drift past the telescope, while collecting images,
produces a scan along the HA axis. We henceforth refer to these scans
as ``HA scans''.

We also made scans in the perpendicular direction, along the
declination angle (DA) axis. For this we used a script that sent
instructions to the telescope pointing program, taking 20\arcsec{} DA
steps while keeping HA zeroed, pausing to let the observer manually
collect a set of 6 images. After making one such DA scan, we decided
that operating the red and blue cameras at the same time while doing
the steps was too tiresome and error prone. We then scanned one CRISP
wavelength at a time (skipping 5576) and the blue cameras separately.
Our intention was to analyze only the latter data but we later
realized that the first scans were useful for checking the consistency
of our measurements, so they appear below as ``extra'' data sets.

We did the usual dark and flat corrections on each image using code
from the CRISPRED data pipeline \citep{delacruz15crispred}. From the
flat fields we also established a mask to use in all later processing,
where bad pixels and vignetted areas near the edges of the detectors
are zeroed. For the Sarnoff cameras, the mask includes the vertical
dividers between areas on the detectors that are read out separately.

\subsection{Air mass}
\label{sec:air-mass}

The amount of air between the telescope and the Sun varies with the
angular distance from zenith. Therefore, so does the fraction of the
light that is removed from the direct sunlight by the atmosphere. For
the HA scans, the telescope is pointing in the same direction during
the entire scan but for the DA scans we need to find out whether this
affects our measurements significantly. The telescope software logs
the pointing, so the zenith angle is readily available.

An estimate of the direct sunlight intensity at a certain height $h$
above sea level is given by \citet{pvcdrom}, taking the varying
properties of the atmosphere at different altitudes into account,
\begin{equation}
  \label{eq:Id}
  I_D(\theta_\text{z},h) \propto (1-ah)\cdot
  0.7^{M_\text{A}^{0.678}} + ah,
\end{equation}
where $a=0.14\ \text{km}^{-1}$ and $h\approx2.4$~km for the SST. The
relative air mass, $M_\text{A}$, is the amount of air between the
telescope and the Sun, normalized to unity at zenith.
\citet{kasten89revised} give an approximation of $M_\text{A}$, taking
the curvature of the earth (and the atmosphere) into account,
\begin{equation}
  \label{eq:3}
  M_\text{A}(\theta_\text{z}) = \bigl(   
  \cos\theta_\text{z} + 0.50572\cdot(96\fdg07995-\theta_\text{z})^{-1.6364}
  \bigr)^{-1},
\end{equation}
where $\theta_\text{z}$ is the angular distance from zenith.

Equations (\ref{eq:Id}) and (\ref{eq:3}) are both semi-empirical,
involving measurements at specific sites. Therefore we have verified
that they are valid for our site by comparison with intensity
measurements from the SST SHABAR\footnote{The shadow band ranger
  (SHABAR) of the SST is one of a set of five, constructed by
  \citet{2010SPIE.7733E..4LS}. A SHABAR measures intensity variations
  (scintillation) in an array of photo diodes pointing toward the Sun.
  From those measurements it estimates the index of refraction
  structure parameter, $C_n^2$, along the line of sight. The average
  of the photo diode intensities is a measure of the direct
  sunlight.}. There is good agreement for $\theta_\text{z} < 70\degr$
(Sliepen 2015, private communication).

With the DA zenith angles in Table~\ref{tab:observations} the
variations in intensity are less than 0.22\% during a single scan.
This is negligible for our purposes so we made no correction for
varying air mass.

\subsection{Pointing and coordinate transformations}
\label{sec:coord-transf}

The Primary Image Guider \citep[PIG;][]{sliepen13primary} of the SST
measures the pointing of the telescope on the solar disk by tracking
the primary image at the bottom of the telescope vacuum tube. In
addition to the solar disk, the instrument can see the field stop of
the vacuum exit window. It can therefore keep track of what part of
the disk is in the science FOV. The PIG logs disk coordinates with a
modest accuracy of $\sim$30\arcsec{} but the precision is much better,
on the order of 1\arcsec. After a calibration of the translation
of the coordinate system, the expected accuracy should be as good as
the precision, definitely better than image shifts from seeing. In
order to properly model the data in 2D, we also need to find out the
rotation angle between the logged coordinates and the coordinate
system given by the pixel axes of the image data frames. Both
calibrations can be done for each scan using images from the limbs.

\begin{figure}[!tb]
  \centering
  \includegraphics[viewport=76 43 700 528,clip,width=\linewidth]{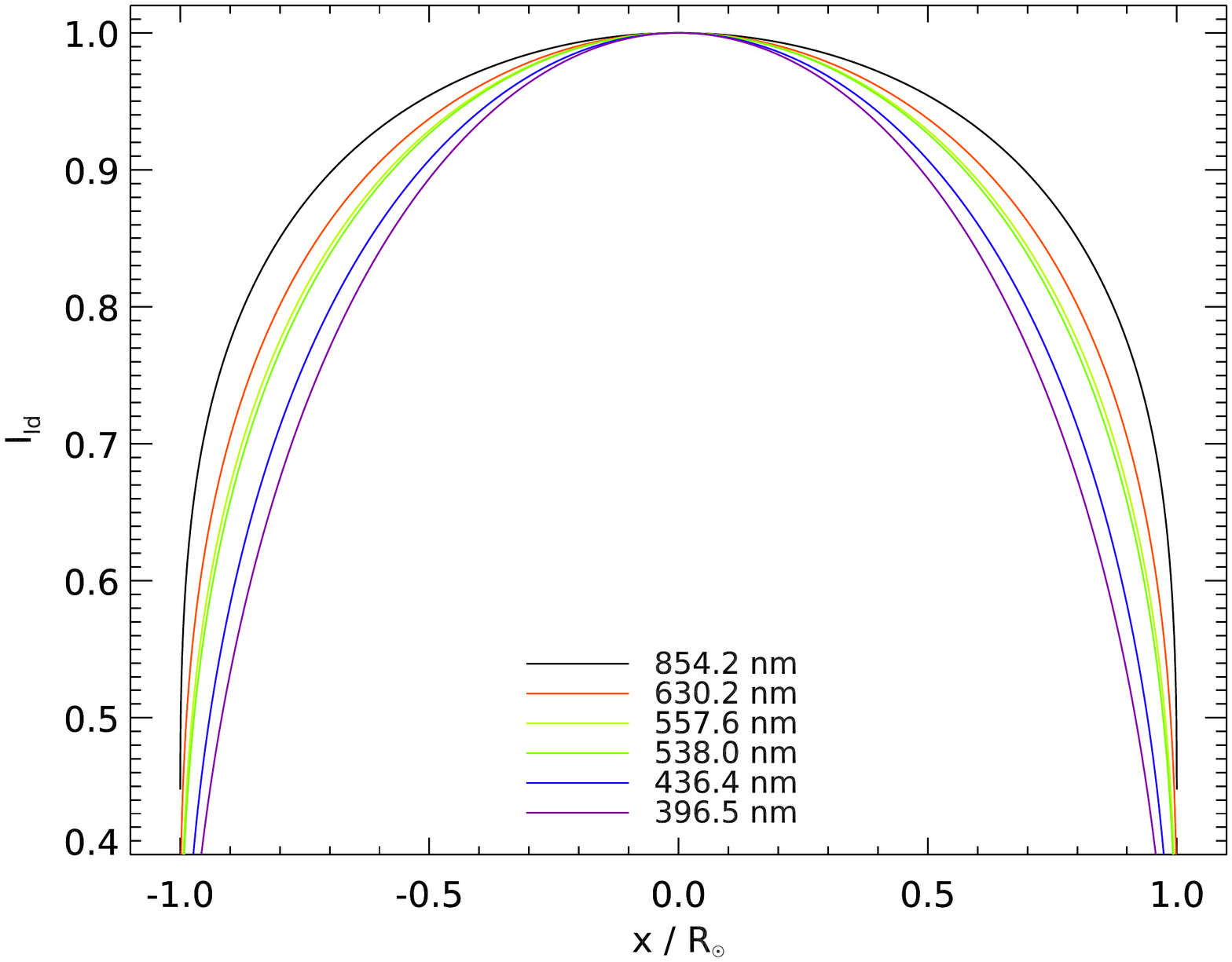}
  \caption{Limb darkening from \citet{1994SoPh..153...91N}.}  
  \label{fig:limb-dark-neckel}
\end{figure}

The procedure is to find the images from each limb, that divide the
FOV as evenly as possible (worked excellently for the HA scans,
decently for the DA scans). We then fit a 2D model with free
parameters including limb distance from FOV center, angle toward DC,
logged limb image distance, blurring kernel width, and a
multiplicative fudge factor (close to unity). For the limb darkening
we used a fifth order polynomial in the cosine of the viewing angle,
$\mu =\cos\bigl(\arcsin(r/R_\sun)\bigr)$, apparently first done by
\citet{1977SoPh...51...25P},
\begin{equation}
  \label{eq:6}
  h(r\,;p_0,p_1,p_2,p_3,p_4,p_5)=\sum_{j=0}^5 p_j\, \mu^j(r) .
\end{equation}
We calculated $R_\sun=953\farcs4$ for 27 April 2014 using code from
the \texttt{get\_sun} subprogram in Solarsoft
\citep{1998SoPh..182..497F}. For the $p_i$ parameters we used values
interpolated in wavelength from the empirical data of \citet[their
Table~1]{1994SoPh..153...91N}. Figure~\ref{fig:limb-dark-neckel} shows
the thus calculated limb darkening curves for our wavelengths.

\begin{figure}[!tb]
  \centering
  \subfloat[Observed]{%
    \begin{minipage}[c]{0.195\linewidth}
      \includegraphics[viewport=0   0 255 255, width=\linewidth, clip]{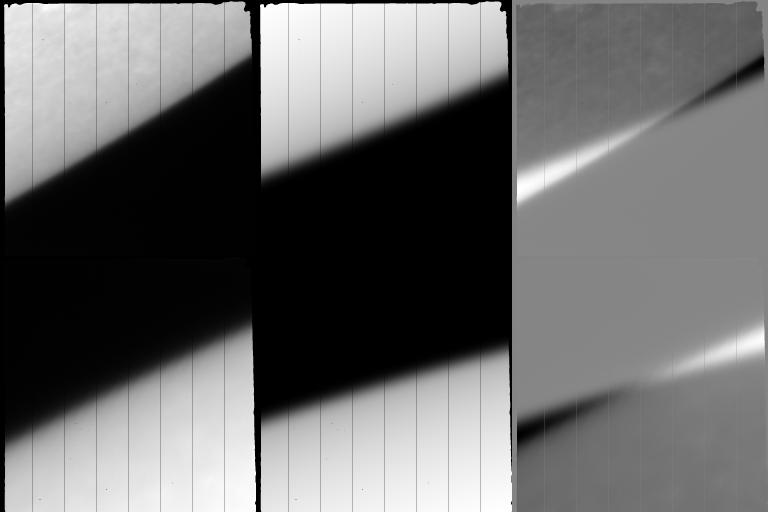}\\[0.2mm]
      \includegraphics[viewport=0 256 255 511, width=\linewidth, clip]{27137fg3a}
    \end{minipage}}
  \hfill
  \subfloat[Initial model]{%
    \begin{minipage}[c]{0.39\linewidth}
      \includegraphics[viewport=256   0 767 255, width=\linewidth, clip]{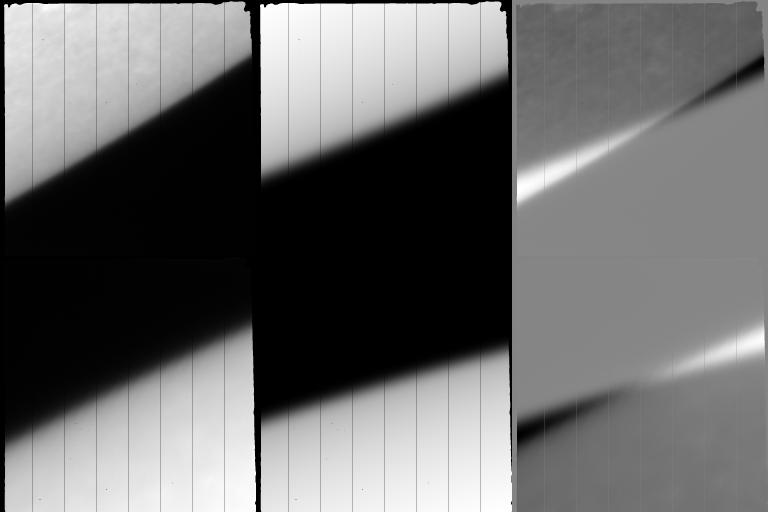}\\[0.2mm]
      \includegraphics[viewport=256 256 767 511, width=\linewidth, clip]{27137fg3b}
    \end{minipage}}
  \hfill
  \subfloat[Converged model]{%
    \begin{minipage}[c]{0.39\linewidth}
      \includegraphics[viewport=256   0 767 255, width=\linewidth, clip]{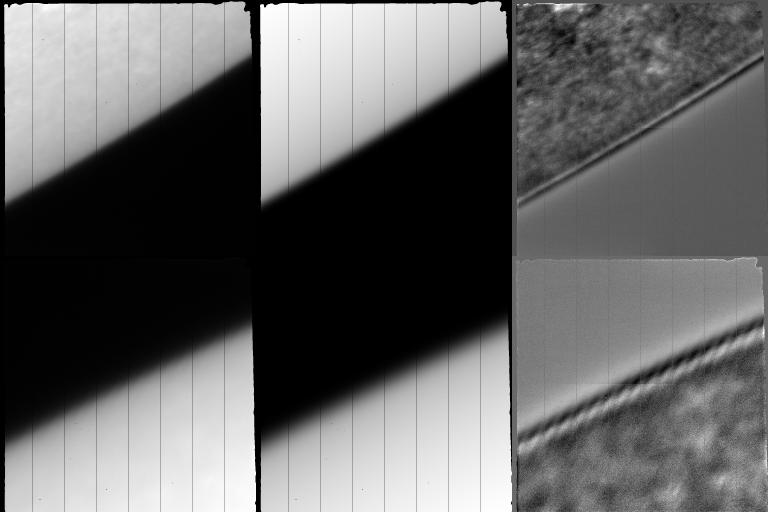}\\[0.2mm]
      \includegraphics[viewport=256 256 767 511, width=\linewidth, clip]{27137fg3c}
    \end{minipage}}
  \caption{Limb fit for the 630~nm HA scan. Within each subfigure, the
    top and bottom tiles represent the masked 1024$\times$1024-pixel
    images of the two limbs, respectively. In \textbf{(b)} and
    \textbf{(c)}, the rightmost tiles represent the differences
    between the model images and the observed images.}
  \label{fig:limbfit}
\end{figure}

For the initial estimates, we determined the position and orientation
of the limb in each of the limb images with the following heuristics:
\begin{enumerate}
\item Threshold a gradient image to get a mask that selects points
  that are on or close to the limb,
\item Find the coordinates of the thus selected limb point that is the
  one closest to the center of the FOV,
\item Crop the image to as large an area as possible, which is
  centered on the selected limb point,
\item Divide the cropped image into four equal parts, forming a quad
  cell where the maximum value indicates the image quadrant nearest to
  DC,
\item Calculate the angle to DC within the quadrant as the arctangent
  of the two quad cell values surrounding the maximum cell.
\end{enumerate}
This procedure gave good enough initial estimates for the model fit to
converge nicely. For sample model fitting results, see
Fig.~\ref{fig:limbfit}. Evidently, the Neckel \& Labs limb darkening
curves were good enough for these fits. However, as we will see in the
next section, they are not perfect fits to our data.

We performed these fits (as well as the fits in the following
sections) in IDL with the MPFIT package
\citep{more78levenberg,2009ASPC..411..251M}.

\subsection{Binning}
\label{sec:binning}

For each image in a scan, we calculated the radial coordinate of each
pixel-based on the coordinates logged by PIG and the transformation
established from the limb images (see Section~\ref{sec:coord-transf}).
This allowed us to radially bin the intensity data without being
limited by the detector size, and then calculate the average intensity
in each bin. We used a bin size of 10\arcsec. 

This resulted in approximately $3\times10^7$ contributing pixel values
per bin in both red and blue cameras for the HA scans. For the less
densely sampled DA scan data, we got $3\times10^6$ (red) and
$1.3\times10^7$ (blue) pixel values per bin. 

\begin{figure*}[!tb]
  \centering
  \subfloat[HA scans, linear, on disk.\label{fig:limb-dark-data-ha}]{\includegraphics[viewport=76 44 700 528,
    clip,width=0.48\linewidth]{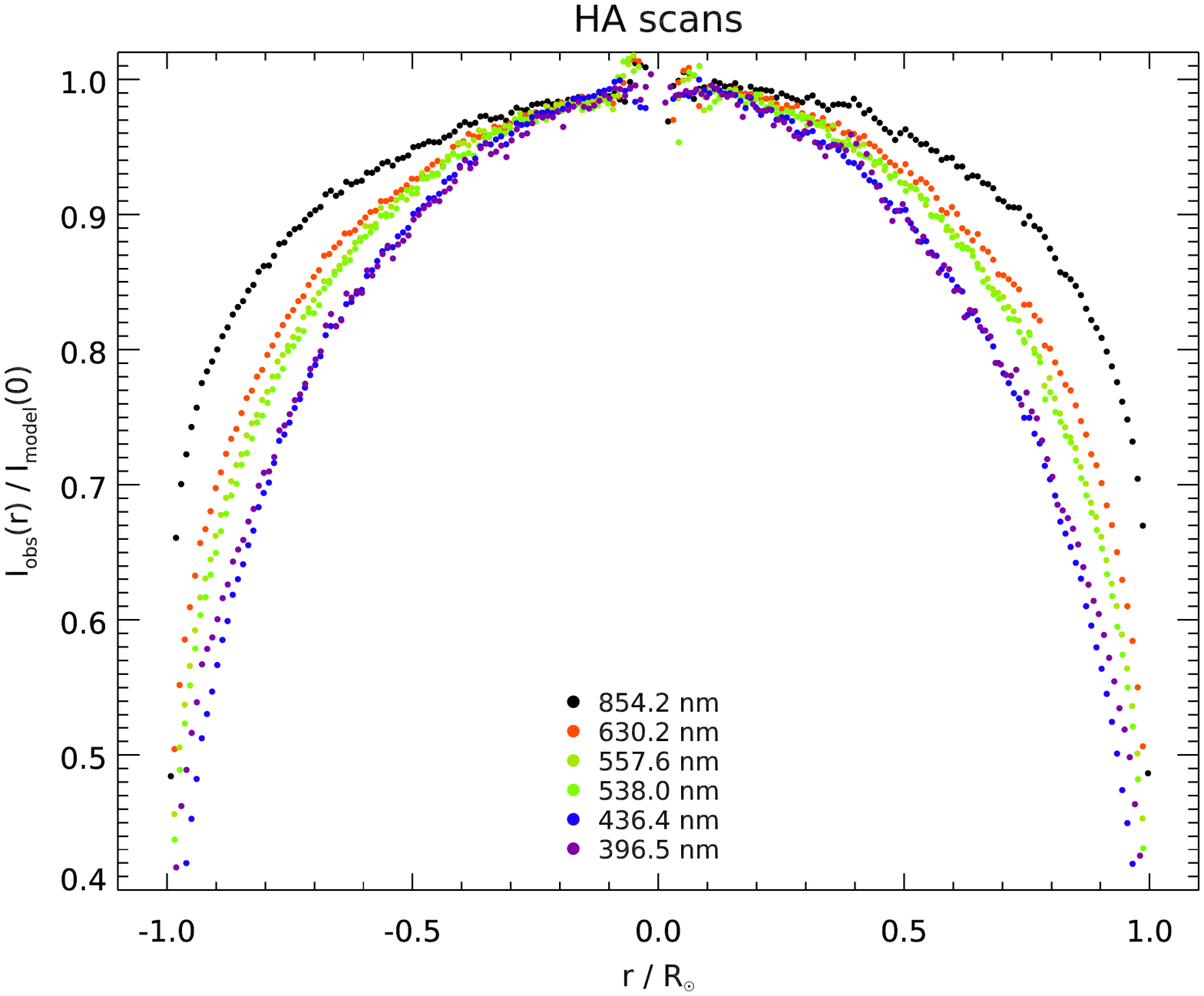}}\hfill
  \subfloat[DA scans, linear, on disk.\label{fig:limb-dark-data-da}]{\includegraphics[viewport=76 44 700 528,
    clip,width=0.48\linewidth]{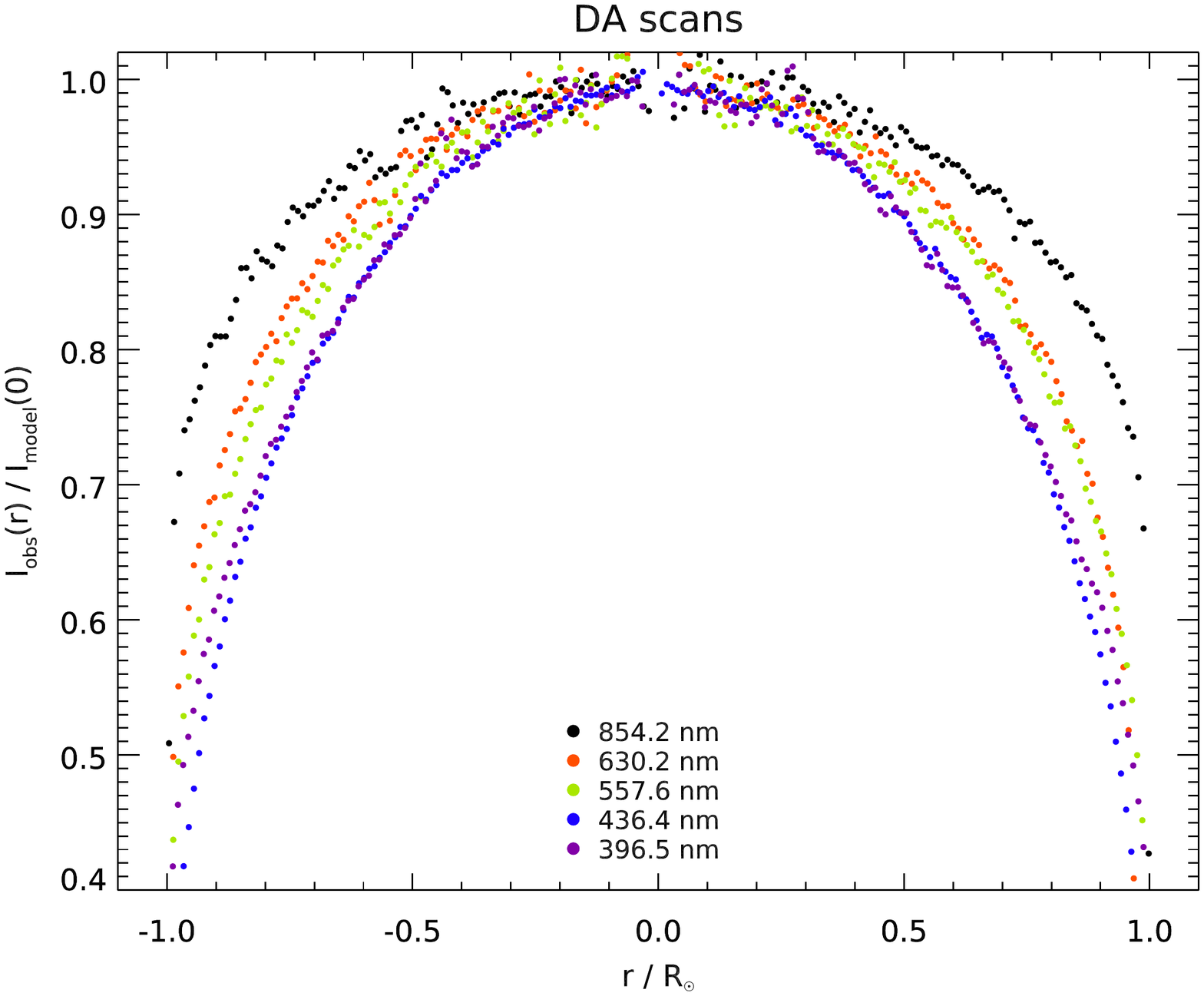}}\\
  \subfloat[HA scans, logarithmic.\label{fig:binned-data-ha}]{\includegraphics[viewport=76 44 700 528,
    clip,width=0.48\linewidth]{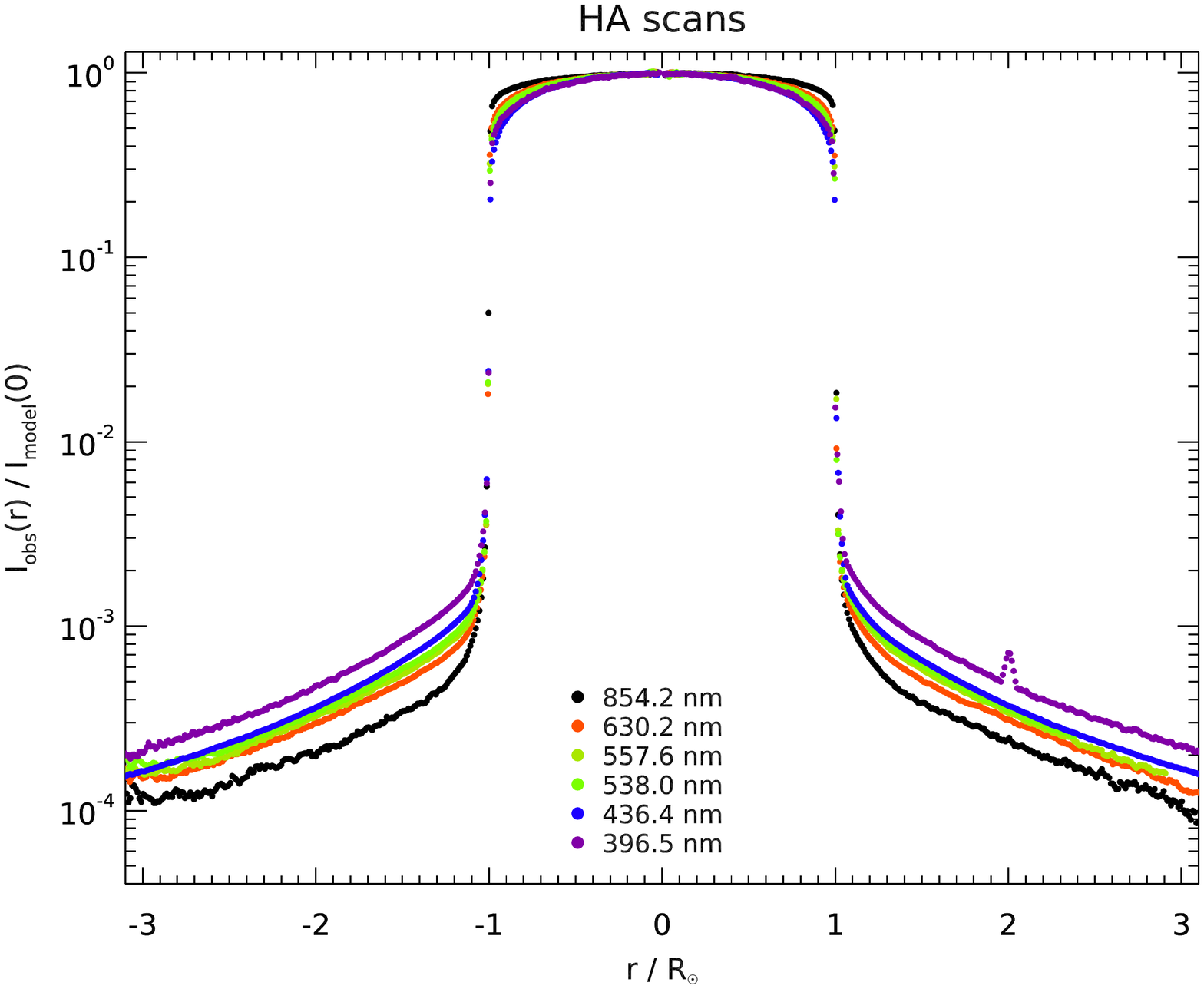}}\hfill
  \subfloat[DA scans, logarithmic.\label{fig:binned-data-da}]{\includegraphics[viewport=76 44 700 528,
    clip,width=0.48\linewidth]{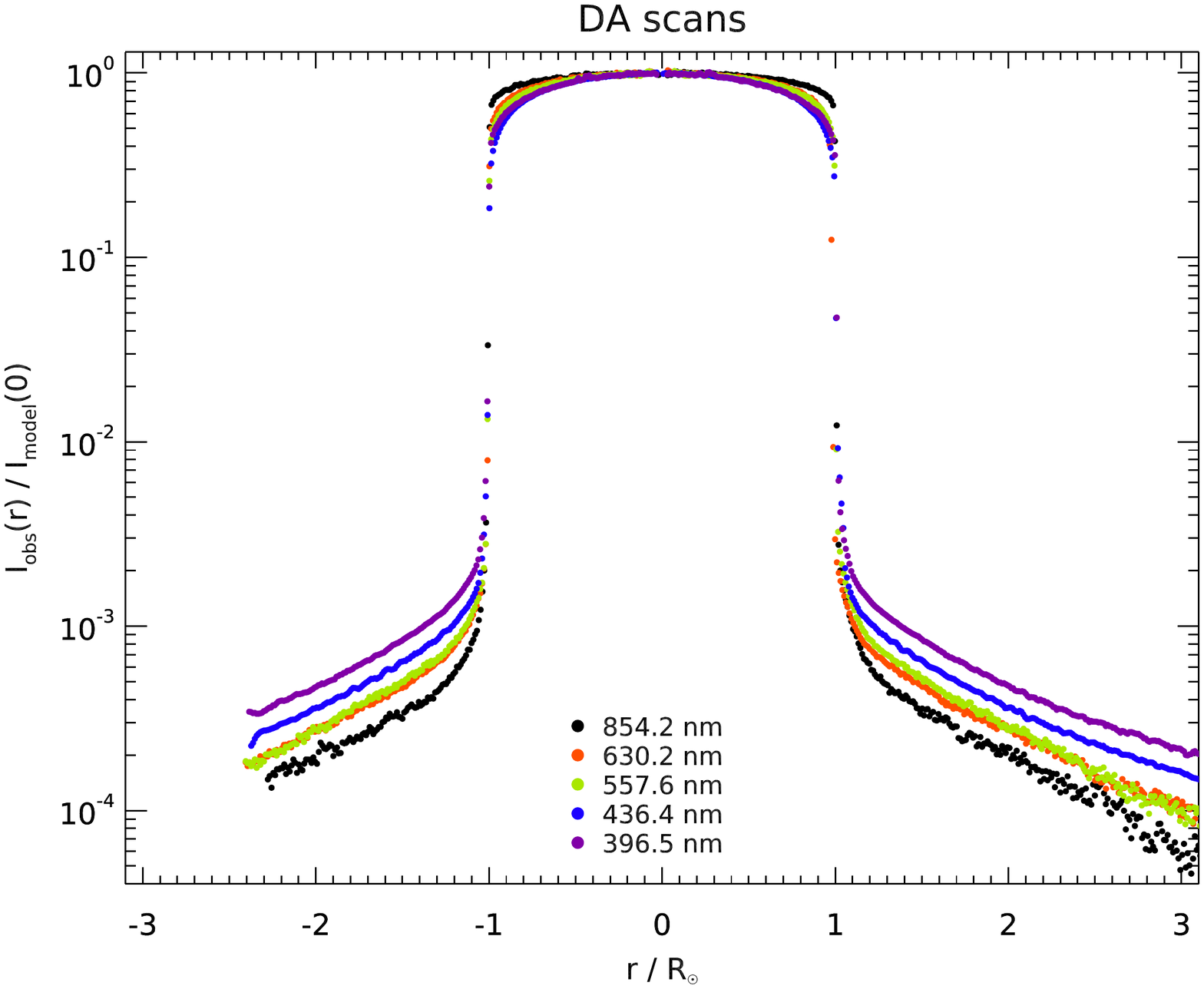}}
%  % plotdata.pro
  \caption{Binned and normalized intensity data (10\arcsec{} bins) of
    the HA and DA scans. Compare Fig.~\ref{fig:limb-dark-neckel}.
    Positive (negative) $r$ correspond to scan direction away from
    (toward) DC. The 11:49 ``extra'' DA are scans omitted from these
    plots.}
  \label{fig:binned}
\end{figure*}

The binned intensity data are plotted in Fig.~\ref{fig:binned},
normalized to DC intensity as given by the fits described in
Section~\ref{sec:straylight-analysis} below. In this figure, we have
plotted the toward-DC halves (TDHs) of the scans with negative radial
coordinates, while the from-DC halves (FDHs) are plotted with positive
coordinates. The linear plots show the limb darkening for the
different wavelengths, while the logarithmic plots show the faint
off-disk straylight, demonstrating that there is signal all the way
out to $3R_\sun$.

Due to the smaller number of pixel values per bin, the CRISP DA scan
data are noticeably noisier on the disk than the HA scan data (compare
Figs.~\ref{fig:limb-dark-data-ha} to~\ref{fig:limb-dark-data-da}
and~\ref{fig:binned-data-ha} to~\ref{fig:binned-data-da}).
For both types of scan, the points within $0.1R_\sun$ from DC are
noisier because our scans did not go exactly through DC so for small
$r$ we get contributions from just a few pixels in the corners of
image frames or none at all.

The \citet{1994SoPh..153...91N} limb darkening curves fit several of
our scans fairly well, as can be seen by comparing
Fig.~\ref{fig:binned} with Fig.~\ref{fig:limb-dark-neckel}. However,
the 396~nm data deviate from the regular progression with wavelength
and the 854~nm data do not fall off with $r$ quite as quickly as in
the Neckel \& Labs curve. We conjecture that the reason in both cases
is that the filter passbands are completely within wide \ion{Ca}{ii}
lines. In the 396~nm case in the wing of the K line and near some deep
blends and in the 854~nm case including the chromospheric line core.
The Neckel \& Labs measurements were done for continuum wavelengths.

\subsection{The aureola intensity}
\label{sec:off-disk-intensity}

\begin{figure*}[p]
  \centering
  \def\tilewidth{0.495\linewidth}
  \subfloat[396.5 nm]{\includegraphics[viewport=50 44 700 528,clip,width=\tilewidth]{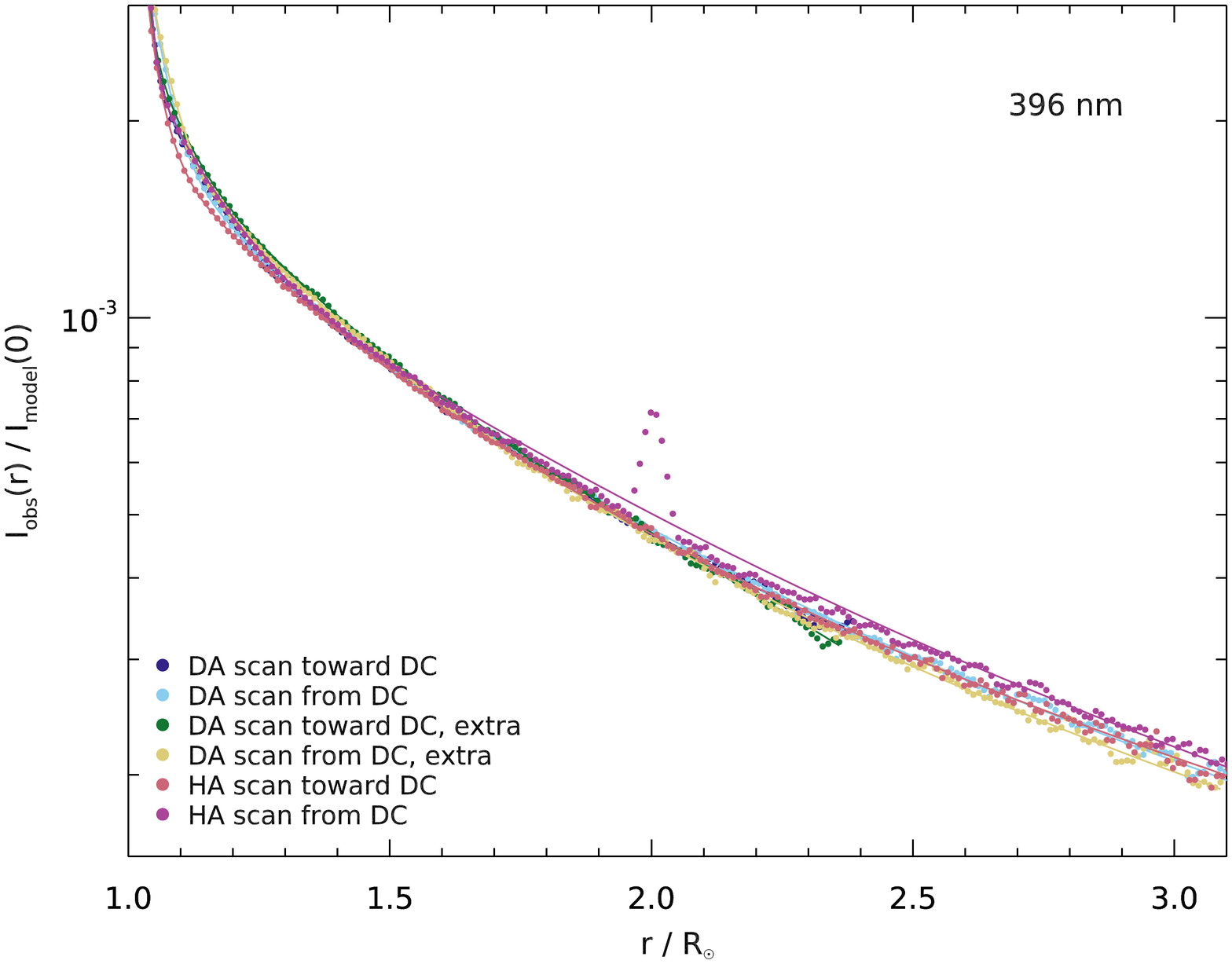}}\hfill
  \subfloat[436.4 nm]{\includegraphics[viewport=50 44 700 528,clip,width=\tilewidth]{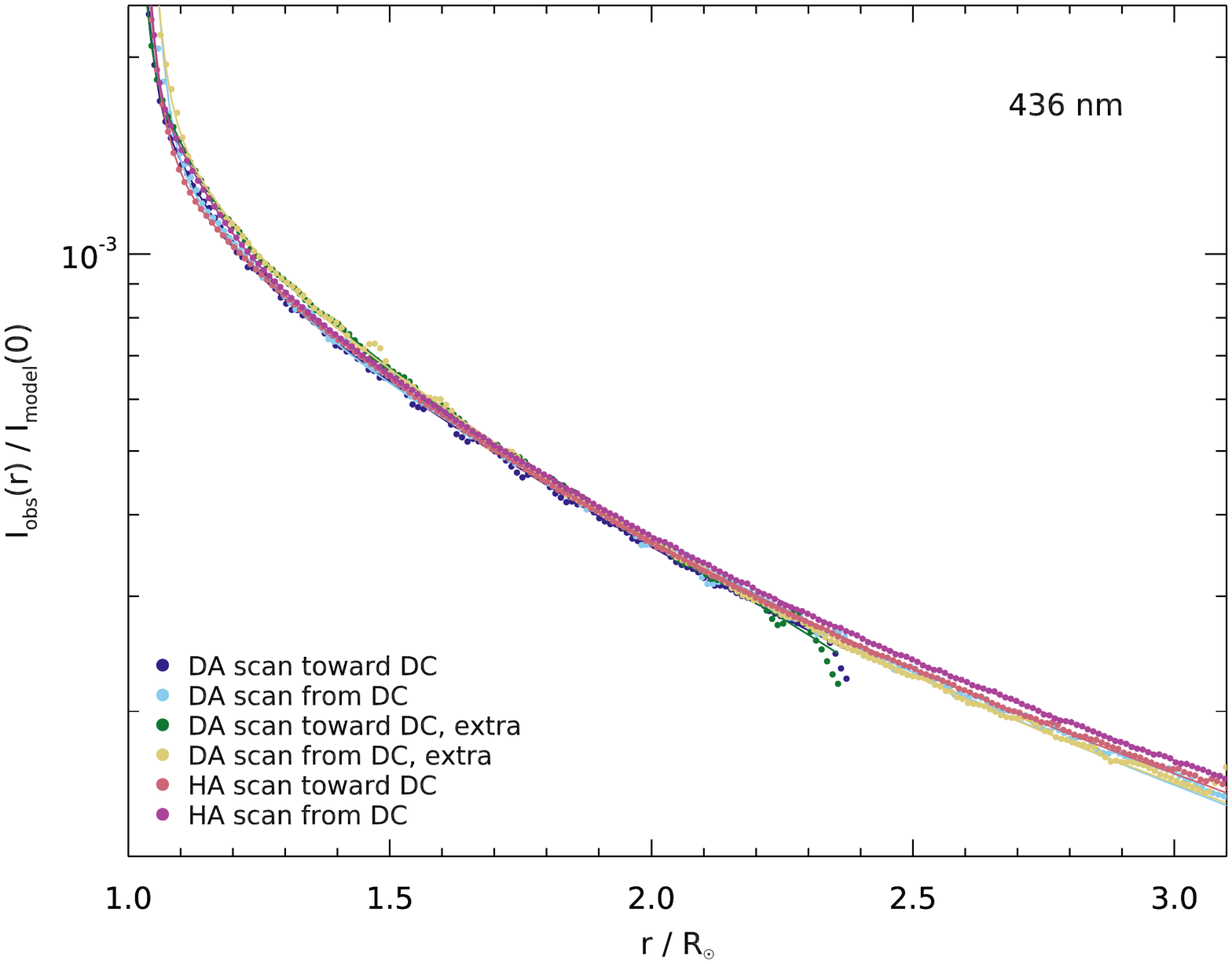}}\\
  \subfloat[538.0 nm]{\includegraphics[viewport=50 44 700 528,clip,width=\tilewidth]{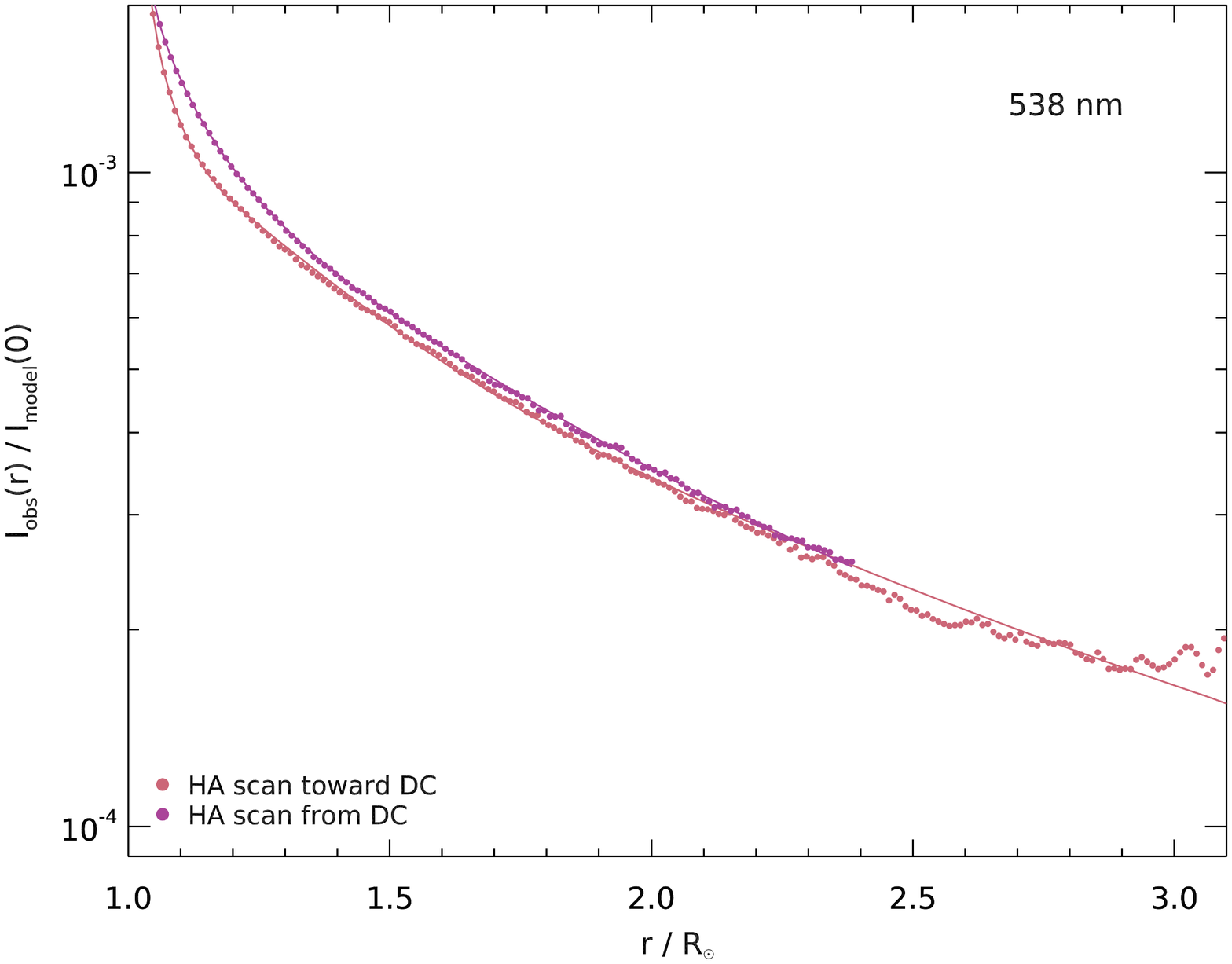}}\hfill
  \subfloat[557.6 nm]{\includegraphics[viewport=50 44 700 528,clip,width=\tilewidth]{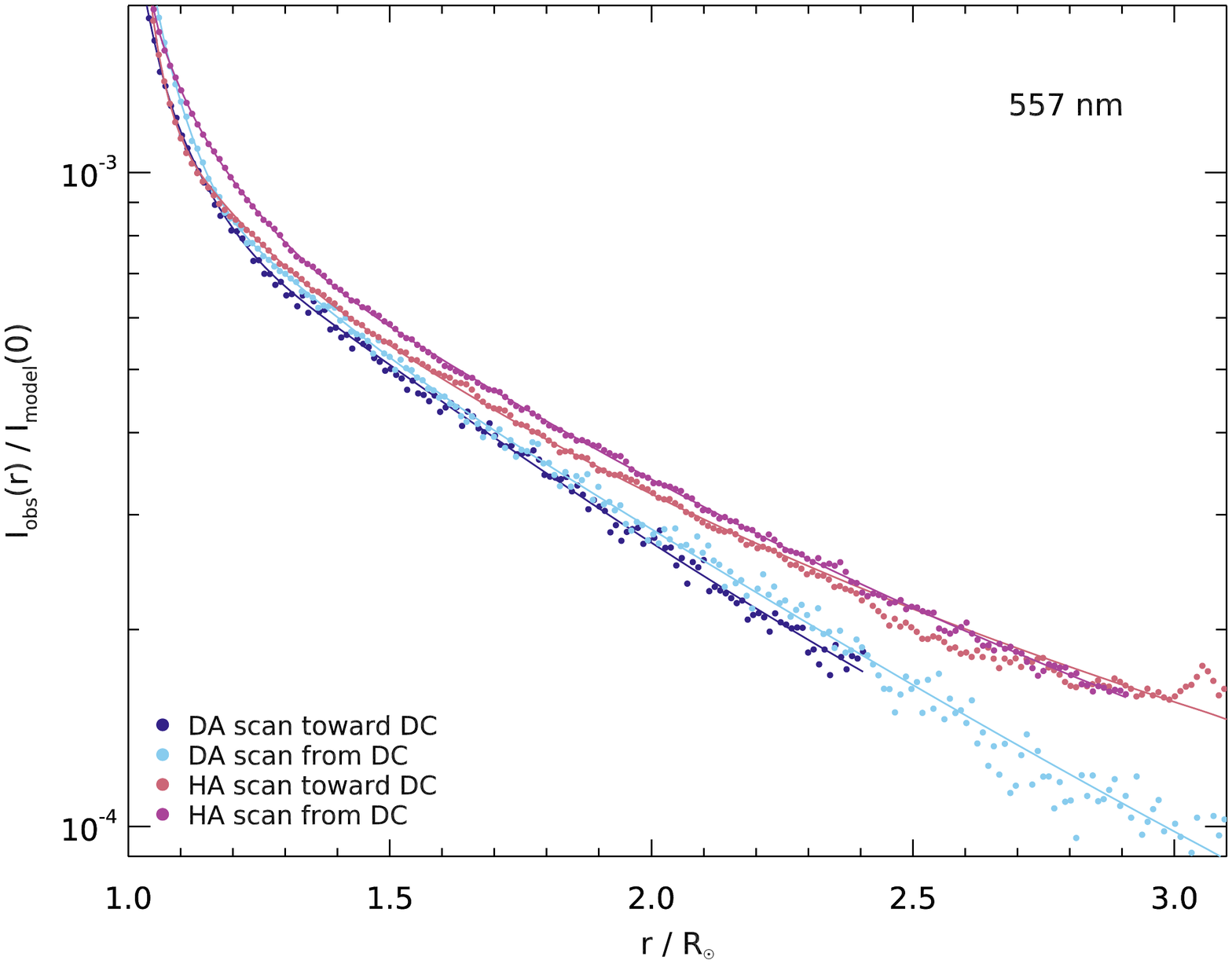}}\\
  \subfloat[630.2 nm]{\includegraphics[viewport=50 44 700 528,clip,width=\tilewidth]{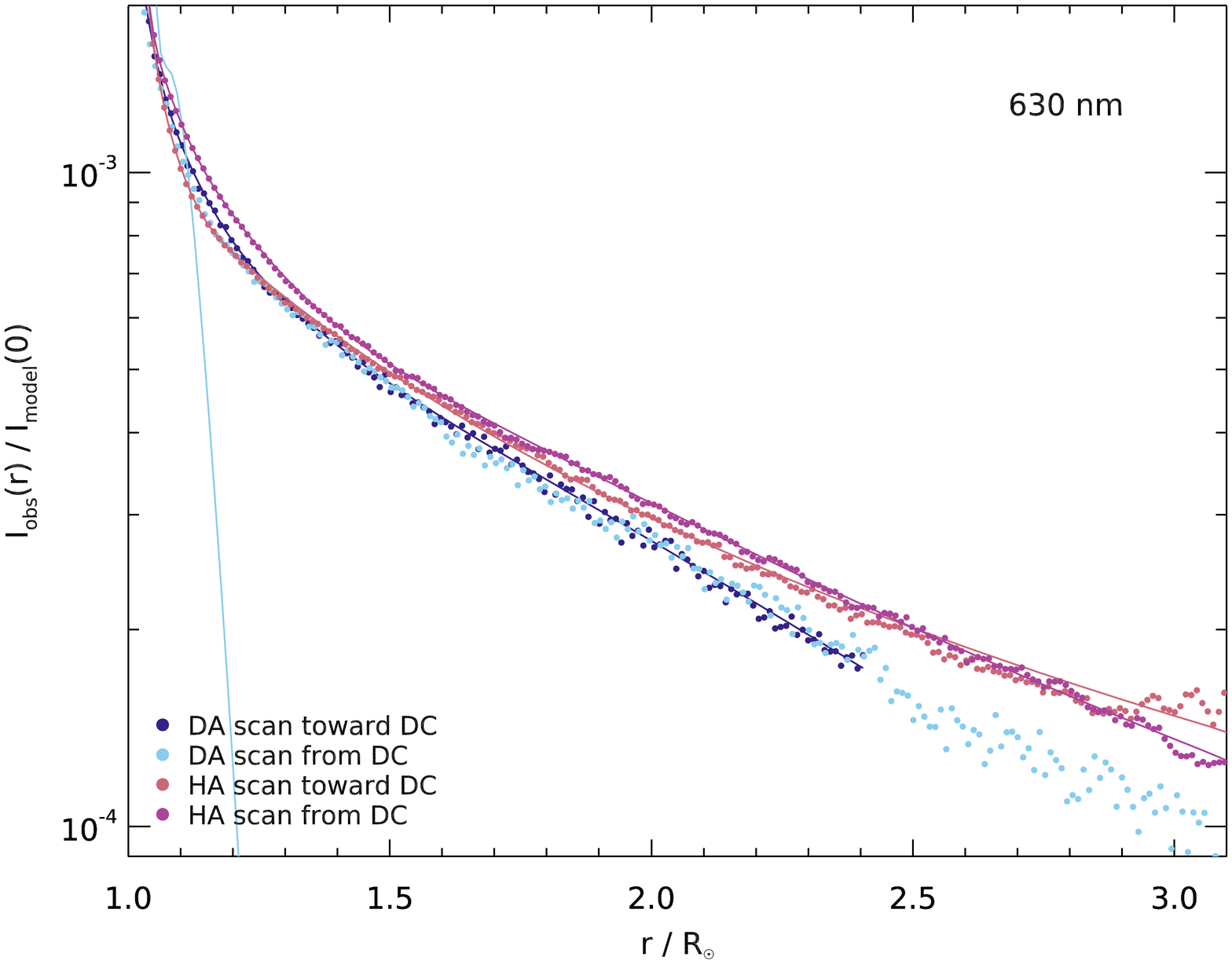}}\hfill
  \subfloat[854.2 nm]{\includegraphics[viewport=50 44 700 528,clip,width=\tilewidth]{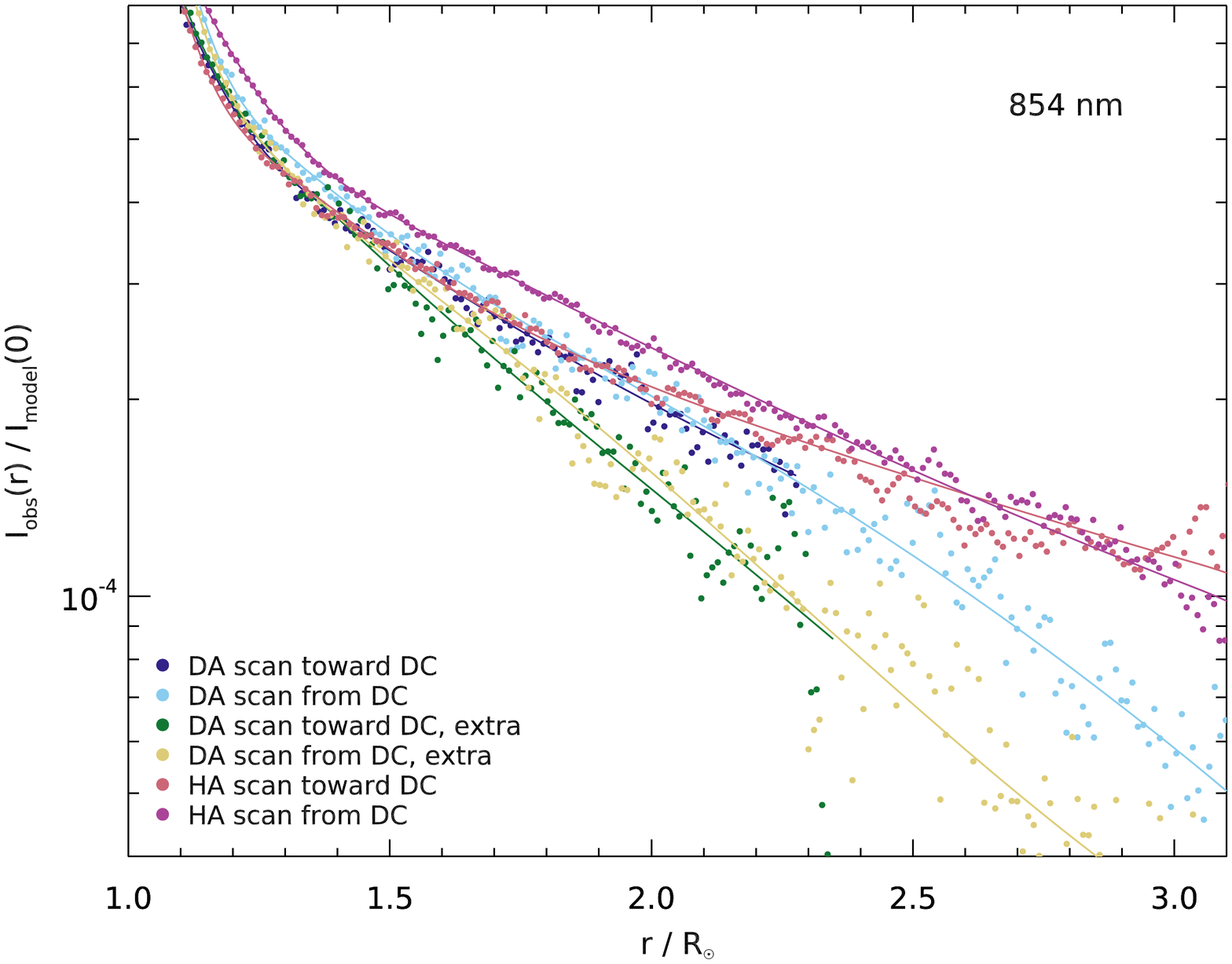}}\\
  \caption{Aureola intensity of CRISP WB and blue scans. The dots
    represent the binned data, while the lines represent the fits
    described in Section~\ref{sec:straylight-analysis}, specifically
    the ones in Table~\ref{tab:total-all}. Normalization and $\delta
    r$ from the fits. Note that the range on the vertical axis varies
    between the different plots.}
  \label{fig:wings}
\end{figure*}

The most interesting part of our scans are outside the limb, where the
straylight forms an aureola. We plot them logarithmically in
Fig.~\ref{fig:wings}.

\subsubsection{Deficiencies}
\label{sec:deficiencies}

We first address some deficiencies in the data. There is a leveling
off at $r > 2.5R_\sun$ for the TDHs of the CRISP HA scans. This cannot
be noise, since we do not see the same effect for $r > 2.5R_\sun$. It
is also not seen in the blue data, so it must be originating in the
red part of the beam, most likely vignetting or reflections within the
CRISP optics.
There is a glitch in the FDH of the 396 nm HA scan at $r\approx
2R_\sun$, a premature ending of the FDH of the 538 nm HA scan at
$r\approx 2.4$, and all the TDHs of the DA scans starting at $r\approx
2.5R_\sun$ rather than $3R_\sun$.

\subsubsection{Consistency}
\label{sec:consistency}

For the two blue wavelengths, 396 and 436~nm, there is a very good
consistency in the off-disk intensity. Not only between the two halves
of the scans, but also between scans collected at different times and
between the orthogonal HA and DA scans.

For all the CRISP wavelengths, the FDHs of the HA scans show a
near-limb ($r<1.4R_\sun$) straylight component that is wider than that
of the TDH. The far FDH components are also slightly (for 854 nm:
significantly) higher than those of the TDHs. The far components of the
DA scans are significantly lower than both halves of the HA scans,
particularly so for the ``extra'' 854 nm scan. 

For the two blue filters, the scans collected with different zenith
distances are remarkably consistent, indicating an instrumental
origin. For the three CRISP prefilters with both HA and DA scans, the
DA scans are collected closer to zenith and also have less straylight,
consistent with a significant atmospheric contribution that correlates
with air mass. The average zenith distance of the 854~nm ``extra
scan'' collected around 12~UT is approximately the same as for the DA
scan collected after 14~UT (local noon in La Palma is a few minutes
past 13~UT) and yet it shows even less intensity in the aureola. This
is consistent with a decrease in the dust concentration with time and
an insignificant atmospheric contribution to the blue scattered light.
However, with a very sparse $\theta_\text{z}$ sampling and lacking a
detailed record of the dust concentration, we do not believe our data
support a detailed analysis with the aim of properly separating the
atmospheric and instrumental contributions.

\subsubsection{The CRISP FPI optics}
\label{sec:crisp-nb}

In addition to the WB data discussed so far, with CRISP we
simultaneously collected continuum NB data. As the WB data are
dominated by the continuum wavelengths, this allows us to see whether
any significant straylight comes from the FPI and LC optics of CRISP.

\begin{figure*}[tp]
  \centering
  \def\tilewidth{0.495\linewidth}
  \subfloat[538.0 nm]{\includegraphics[viewport=50 44 700 528,clip,width=\tilewidth]{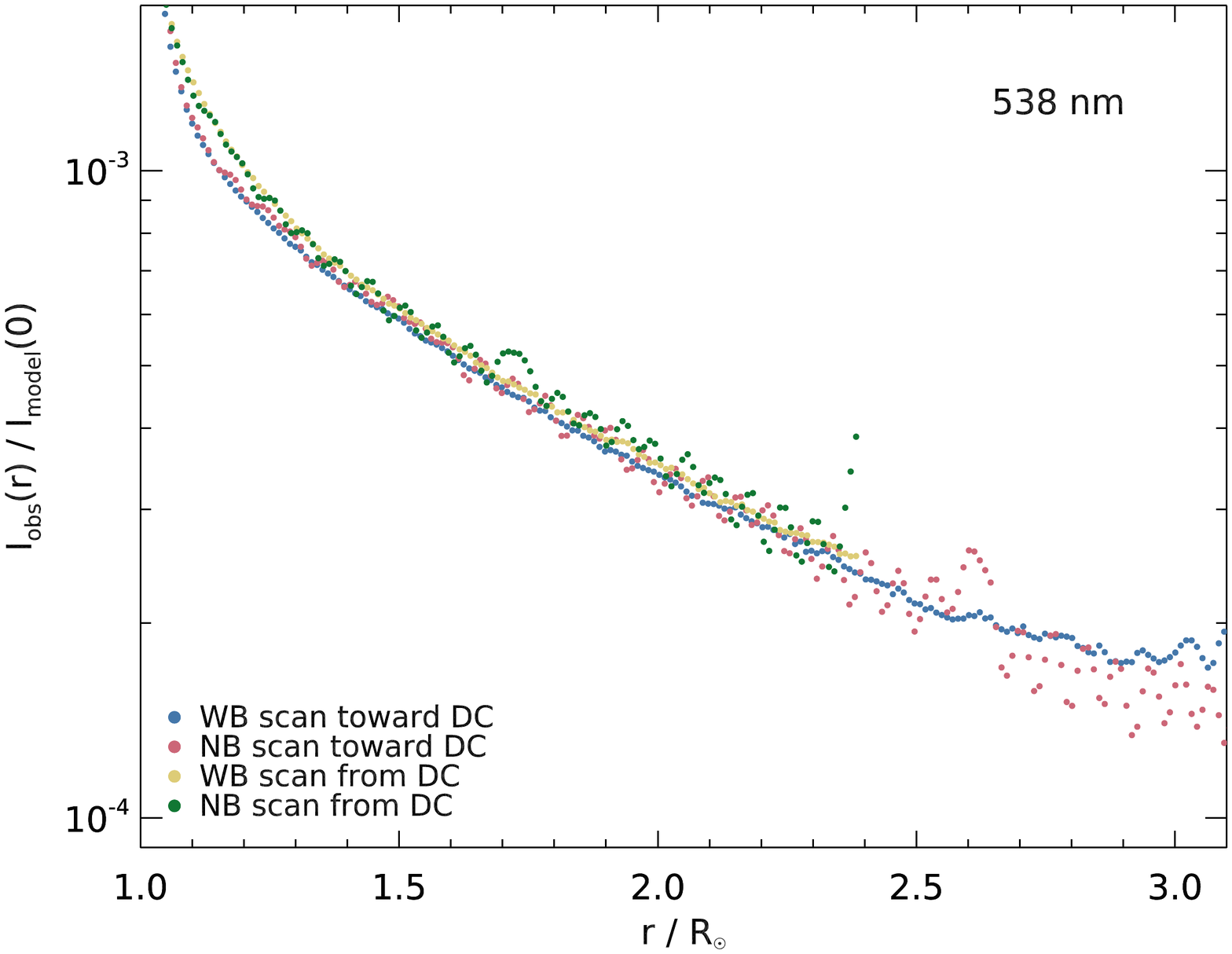}}\hfill
  \subfloat[557.6 nm]{\includegraphics[viewport=50 44 700 528,clip,width=\tilewidth]{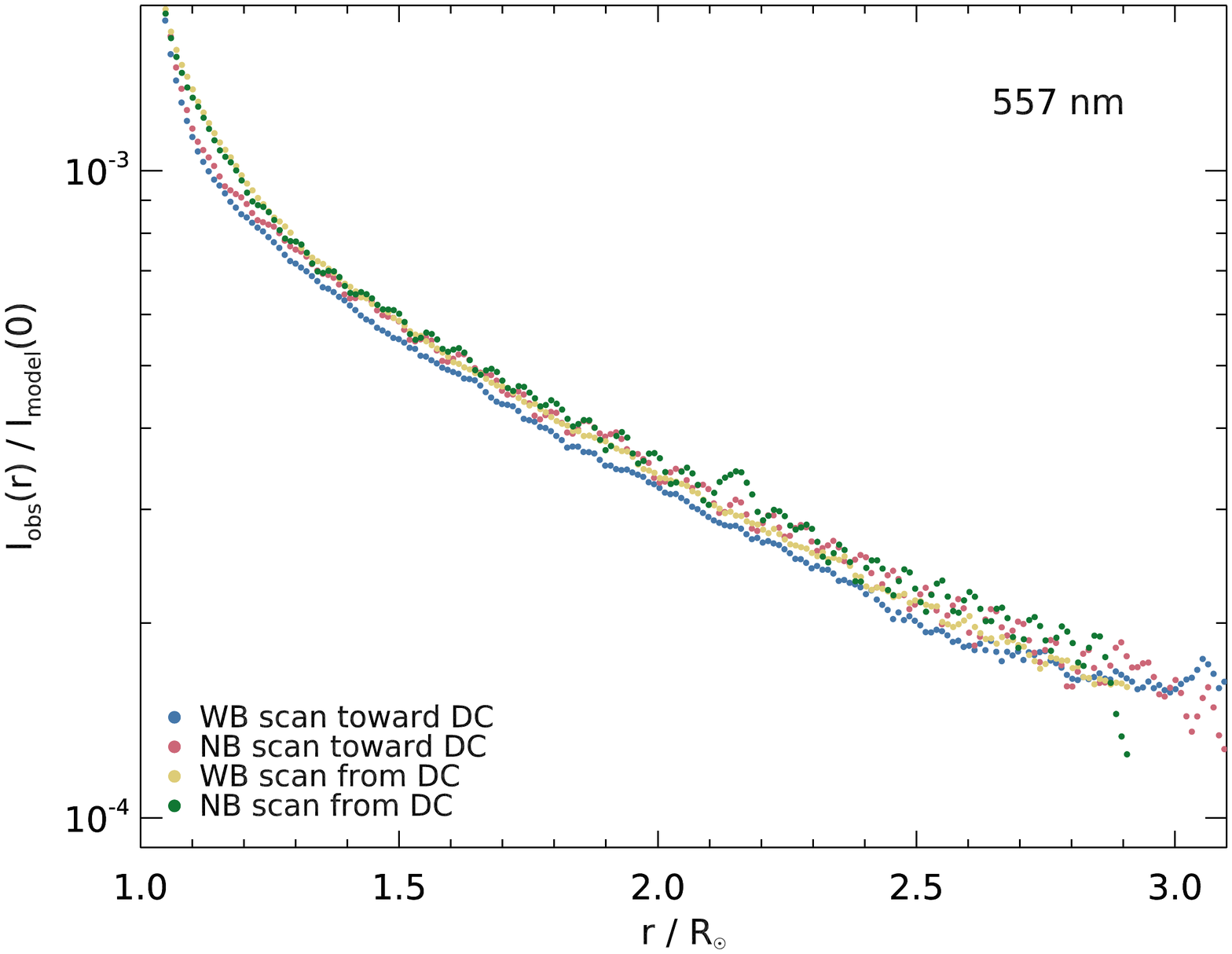}}\\
  \subfloat[630.2 nm]{\includegraphics[viewport=50 44 700 528,clip,width=\tilewidth]{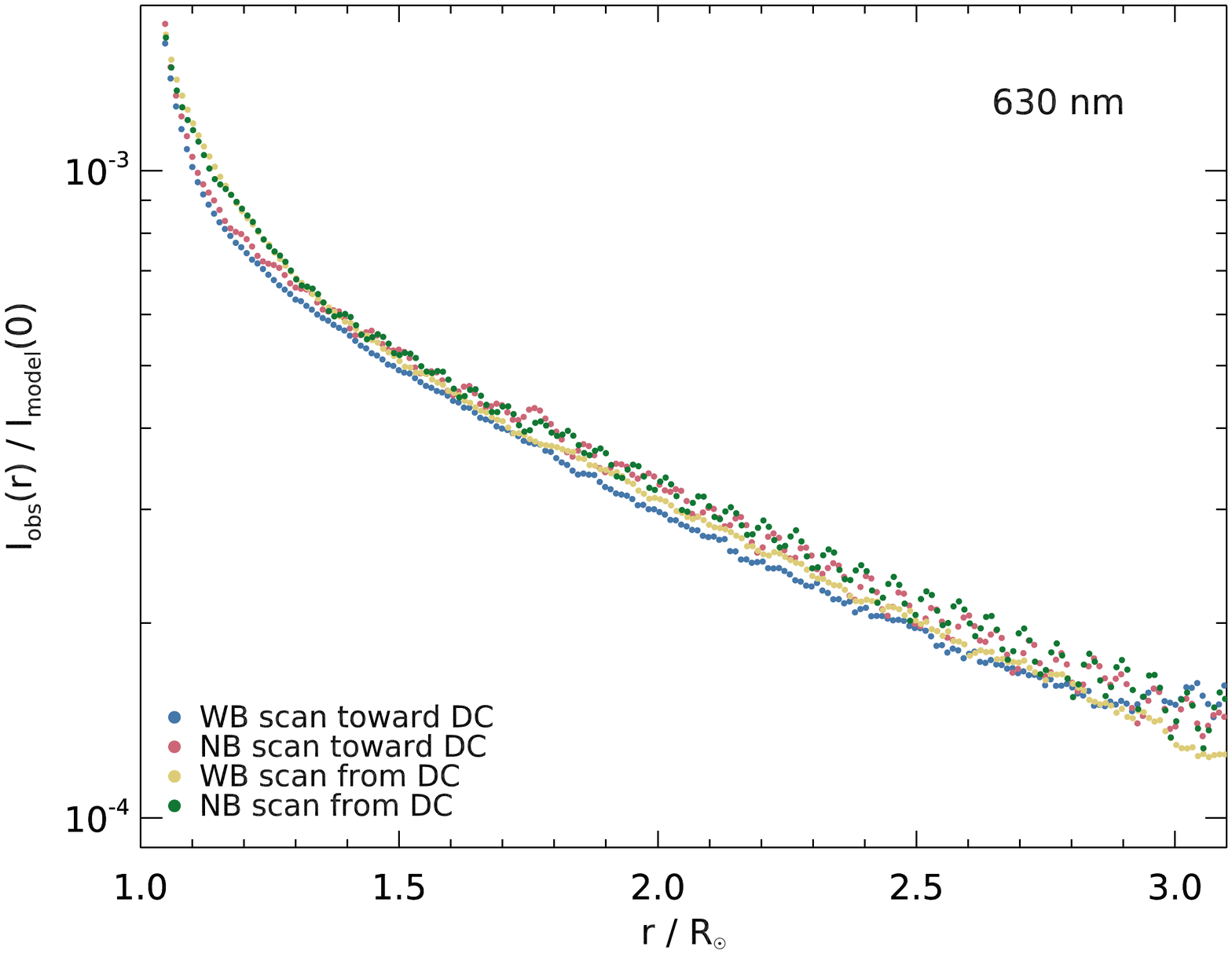}}\hfill
  \subfloat[854.2 nm]{\includegraphics[viewport=50 44 700 528,clip,width=\tilewidth]{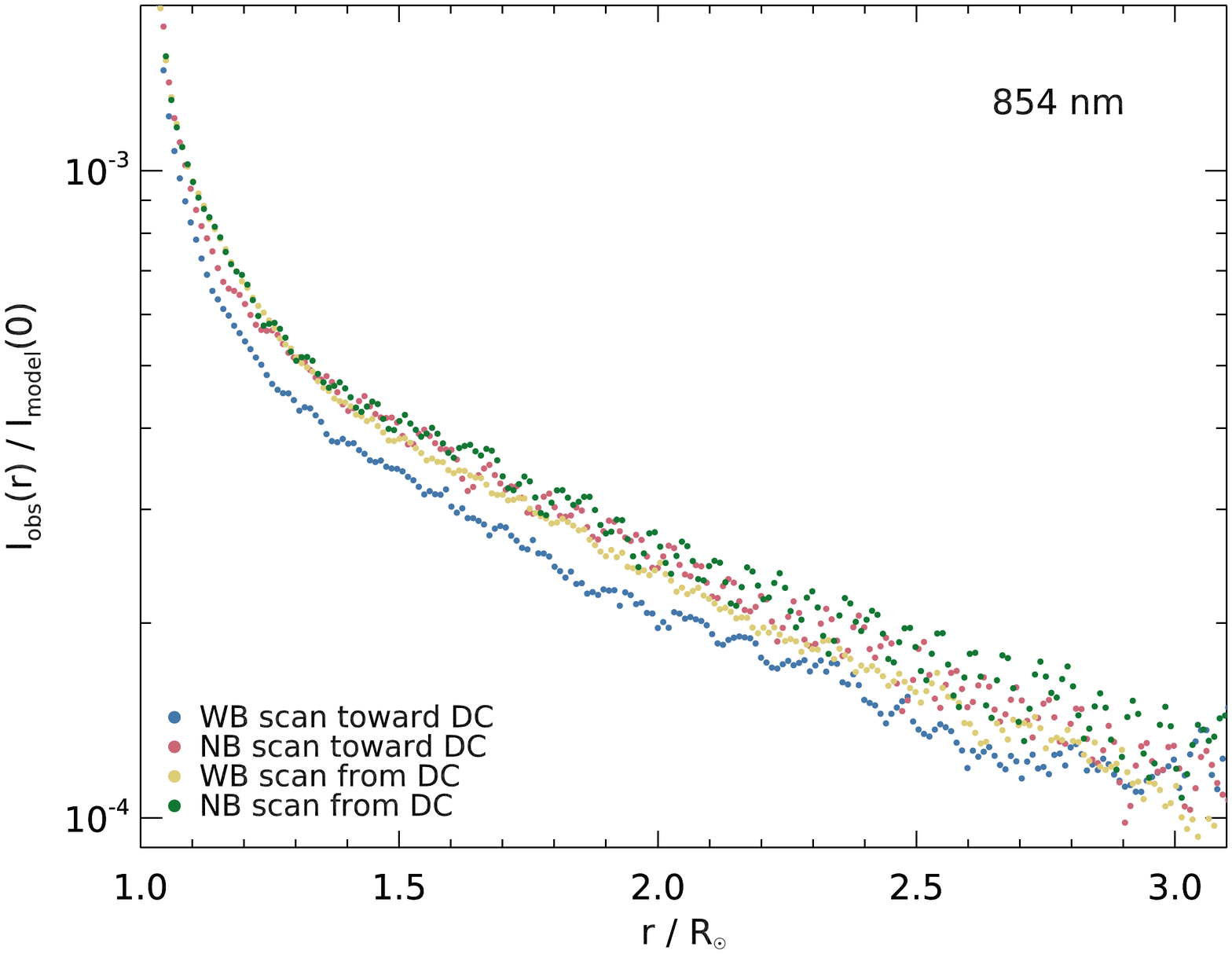}}\\
  \caption{Consistency in HA scan aureola intensity, CRISP NB vs WB.
    Normalization and $\delta r$ from the 3-Gauss fit to the WB data.}
  \label{fig:wingsNB}
\end{figure*}

In Fig.~\ref{fig:wingsNB} we compare the WB and NB continuum HA scan
measurements. The WB data were normalized by the fits as above. The NB
intensities were scaled to match the WB intensity near DC (but
excluding the noisiest bins) to within a few percent, good enough for
the following comparison. The difference between the aureola
intensities recorded in the NB and the WB are minor compared to the
noise in the NB data, as well as to the difference between the two
halves of the WB scan, for the 538.0~nm, 557.6~nm, and 630.2~nm scans.
A systematic difference at large $r$ can be seen in the 854.2~nm scans
but this is not necessarily due to an increase in straylight, as one
could expect differences in the limb darkening between the NB
continuum and the WB that is dominated by the chromospheric Ca~II
line.

\section{The straylight PSF}
\label{sec:straylight-analysis}

\subsection{Forward model}
\label{sec:forward-model}

The 2D model involves a solar disk with limb darkening, convolution
with a seeing PSF and 2--4 contributions from scattering with
different widths, along with some other parameters. The model
parameters are summarized in Table~\ref{tab:model-parameters}, along
with their limits and initial values used in the fits. Below we
describe the model in more detail.

\begin{table*}[tp]
  \centering
  \caption{Model parameters.}
  \label{tab:model-parameters}
  \begin{tabular}{llllcc@{\,}c@{\,}c@{\,}c}
    \hline\hline\noalign{\smallskip}
    \multirow{2}{*}{Parameter} & 
    \multirow{2}{*}{Description} & 
    \multirow{2}{*}{Initial value} & 
    \multirow{2}{*}{Limit} && 
    \multicolumn{4}{c}{Fit \#}\\
    \cline{6-9}\noalign{\smallskip}
    &&&&& 1 & 2 & 3 & 4 \\
    \hline\noalign{\smallskip}
    $p_0,\ldots,p_5$                            & Limb darkening coefficients                              & From \citet{1994SoPh..153...91N} &            & 
    & \checkmark                                               & \checkmark                       &            & \checkmark                                           \\
    $\delta r$                                  & $r$ coordinate adjustment                                & $0\farcs5$                       &            &  &            & \checkmark &            & \checkmark \\
    $\textsc{fwhm}_\text{S}$                    & width of seeing Gaussian                                  & 5\arcsec                         & $>0$       &  &            & \checkmark &            & \checkmark \\
    $c_1$                                       & 1st NGK weight                    & $0.33\%$                         & $>10^{-5}$ &  &            &            & \checkmark & \checkmark \\
    $\textsc{fwhm}_1$                           & 1st NGK width                     & 57\arcsec                        &            &  &            &            & \checkmark & \checkmark \\
    $c_2$                                       & 2nd NGK weight                    & $0.33\%$                         & $>10^{-5}$ &  &            &            & \checkmark & \checkmark \\
    $\textsc{fwhm}_2$                           & 2nd NGK width                     & 114\arcsec                       &            &  &            &            & \checkmark & \checkmark \\
    $c_3$                                       & 3rd NGK weight                    & $0.33\%$                         & $>10^{-5}$ &  &            &            & \checkmark & \checkmark \\
    $\textsc{fwhm}_3$                           & 3rd NGK width                     & 667\arcsec                       &            &  &            &            & \checkmark & \checkmark \\
    $c_0$                                       & FWK weight                                  & 1.0\%                             & $>0$       &  &            &            & \checkmark & \checkmark \\
    $\alpha$, $\beta$, $\gamma$, $\sigma$       & FWK parameters, see Table~\ref{tab:kernels} & 
    Set to make $\textsc{fwhm}_0\approx R_\sun$ & $>0$                                                     &                                  &            &  & \checkmark & \checkmark                           \\
    \hline
  \end{tabular}
  \tablefoot{The checks in columns with headings 1--4 indicate
    which parameters were fit (not fixed) in the consecutive fits (see
    Table~\ref{tab:r-ranges}). The far wing kernel (FWK) parameters are
    initialized to make the FWHM equal to $R_\sun\approx 956\farcs4$; the
    Moffat $\beta$ parameter is initialized to unity (making it start as a
    Lorentzian). The table shows the initial values for the near-limb
    Gaussian kernel (NGK) parameters $c_i$ and $\textsc{fwhm}_i$ for the
    $N_\text{Gauss}=3$ fits. For $N_\text{Gauss}=1$: $c_1=1\%$;
    $\textsc{fwhm}_1=97\arcsec$. For $N_\text{Gauss}=2$: $c_1=c_2=0.5\%$;
    $\textsc{fwhm}_1=67\arcsec$; $\textsc{fwhm}_2=381\arcsec$. The widths
    of the NGKs were limited during the 3rd fit to make sure they did not
    degenerate but not during the final 4th fit. The initial value for
    $\delta r$ is meant to tell \texttt{mpfit} the expected order of
    magnitude.}
\end{table*}

The intensity at radial coordinate $r$ is modeled as
\begin{equation}
  \label{eq:5}
  g(r) =
  k_\text{S} * \bigl(
  k_\text{G}(r\,;\sigma_\text{s})*h(r+\delta r\,;p_0,p_1,p_2,p_3,p_4,p_5)
  \bigr),
\end{equation}
where $*$ represents convolution, $k_\text{S}$ is a scattering kernel,
and $k_\text{G}(r\,;\sigma_\text{s})$ is a Gaussian kernel
representing seeing with the standard deviation $\sigma_\text{s}$ as a
free parameter. The intrinsic radial intensity distribution of the
disk, $h$, includes a correction term, $\delta r$, in the radial
coordinate. The limb darkening is modeled with the 5th-order
polynomial, $h$, from Eq.~(\ref{eq:6}). We tried using the
coefficients measured by \citet[interpolation in their
Table~1]{1994SoPh..153...91N} but they caused significant errors near
the limb and we suspected that they contributed to the problems we had
with convergence (see Sect.~\ref{sec:fitting-HA} below). As mentioned
above, the bad fits may be because of the non-photospheric
contributions of the spectral lines within the passbands of some of
the filters. We note also that Neckel \& Labs discuss variations in
their scans from scan to scan and from season to season, so their limb
darkening parameters represent some sort of average. Hence, we made
the limb darkening coefficients free model parameters.

We model the scattering kernel as the sum of a Dirac delta function,
$\delta$, and a number of blurring kernels with the relative weights,
$c_i$, as free parameters,
\begin{equation}
  k_\text{S} =  
  \left(
    1 - \sum_{i=0}^N c_i
  \right) 
  \cdot 
  \delta 
  +
  c_0\cdot k_0(r\,;\bullet)
  +
  \sum_{i=1}^{N_\text{Gauss}} c_i\cdot k_G(r\,;\sigma_i)
  .
  \label{eq:2}
\end{equation}
The far wings kernel (FWK), $k_0(r\,;\bullet)$, is one of
$k_\text{M}$, $k_\text{L}$, $k_\text{G}$, or $k_\text{V}$ and
$\bullet$ represents one or two parameters as needed, see
Table~\ref{tab:kernels}. We found, however, that a single kernel would
not produce good fits. We therefore added up to three less wide
near-limb Gaussian kernels (NGKs), $k_\text{G}(r,\sigma_i)$.

\begin{table*}[!tp]
  \centering
  \caption{Far wing kernels}
  \label{tab:kernels}
  \begin{tabular}{llll}
    \hline
    \hline\noalign{\smallskip}
    Kernel name & Definition                                                                        & Full width half maximum                                                                                                                 \\    \hline\noalign{\smallskip}
    Gauss       & $k_\text{G}(r;\sigma)  \propto \exp\bigl(-r^2/(2\sigma^2)\bigr)$                  & $\textsc{fwhm}_\text{G} = \sigma\cdot 2 \sqrt{2 \ln 2}$                                                                                 \\    \noalign{\smallskip}
    Lorentz     & $k_\text{L}(r;\gamma)  \propto (1+r^2/\gamma^2 )^{-1}$                            & $\textsc{fwhm}_\text{L} = \gamma\cdot 2$                                                                                                \\    \noalign{\smallskip}
    Moffat      & $k_\text{M}(r;\alpha,\beta) \propto (1+r^2/\alpha^2)^{-\beta}$                    & $\textsc{fwhm}_\text{M} = \alpha\cdot 2\sqrt{2^{1/\beta}-1}$                                                                            \\\noalign{\smallskip}
    Voigt       & $k_\text{V}(r;\sigma,\gamma)\propto k_\text{G}(r;\sigma) * k_\text{L}(r; \gamma)$ & $\textsc{fwhm}_\mathrm{V}\approx 0.5346\,\textsc{fwhm}_\mathrm{L}+\sqrt{0.2166\,\textsc{fwhm}_\mathrm{L}^2+\textsc{fwhm}_\mathrm{G}^2}$ \\     \hline
  \end{tabular}
  \tablefoot{All kernels are functions of the radial coordinate
    $r=(x^2+y^2)^{1/2}$. They are normalized in the Fourier domain by 
    division with the value in the origin. $k_\text{M}$ is from
    \citet{1969A&A.....3..455M}. The expression for
    $\textsc{fwhm}_\mathrm{V}$ is accurate to within a few
    \textperthousand{} \citep{olivero77empirical}.} 
\end{table*}

%Convolution with FFT requires finite kernels. Although Gaussian wings
%drop fairly quickly to insignificant values, for example Lorentzians
%do not, so it is necessary to truncate the kernels at some radius. How
%wide wings are needed to completely characterize the straylight over
%the entire disk? 
Convolution with FFT requires finite kernels. The wings of Lorentzian
kernels do not drop to insignificant levels quickly, the same is true
for Voigt and Moffat kernels for some parameter values. Therefore it
is necessary to truncate the kernels at some radius, but we need to
keep enough of the wings to completely characterize the straylight
over the entire disk. If we want the light from the entire disk to
potentially spread to every other point on the disk, we obviously need
wings that extend to as far as $2R_\sun$. To allow the entire disk to
contribute to the measurements at $2R_\sun$ outside the disk requires
kernels that are truncated at $r\ga 4R_\sun$. The kernels and the
artificial images used for calculating the convolutions, including
both the solar disk and enough surrounding empty space to protect
against wrap-around contributions from the far wings, were represented
as 2048$\times$2048-pixel arrays.

\subsection{Fitting the model to the scans}
\label{sec:fitting-HA}

We fitted the binned scans separately for the two halves corresponding
to positive and negative~$r$.

Initially, we had problems making the model fit. The limb parameters
$\delta r$ and $\textsc{fwhm}_\text{S}$ tended to diverge before the
limb darkening had converged. A major problem was that the limb
darkening parameters would drive the solution in the wrong direction,
long before the kernels had a chance to reproduce the off-disk
straylight. Before we introduced the NGKs, the FWKs did not converge
well and in particular left large residuals near the limb. But once
introduced, the NGK fits would often degenerate by making $c_i$
go to zero or, for $N_\text{Gauss}>1$, the widths becoming too small
to recover after the limb darkening parameters had converged.

By trial and error, we arrived at the set of parameters described in
the previous section and a procedure that led to convergence (in most
cases, see below). For each data set, we called the \texttt{mpfit}
program four times, with the results from one fit being used as an
initial estimate for the following. The final columns in
Table~\ref{tab:model-parameters} show what parameters were fitted
during each of the four fits, the others were fixed at their initial
values or the values from the previous fit. In each fit, we minimized
$\chi^2$ for different ranges in the radial coordinate, see
Table~\ref{tab:r-ranges}. We used the numbers of contributing pixels
to each bin as weights, this way the noisy bins near DC did not seem
to cause any problems.

\begin{table}[!tbp]
  \centering
  \caption{Ranges in $r$, consecutive fits.}
  \label{tab:r-ranges}
  \begin{tabular}{llll}
    \hline
    \hline\noalign{\smallskip}
    Fit \# & lower limit & upper limit \\
    \hline\noalign{\smallskip}
    1 & 0                    & $R_\sun$              \\
    2 & 0                    & $R_\sun + 30\arcsec $ \\
    3 & $R_\sun + 30\arcsec$ & --                    \\
    4 & 0                    & --                    \\
    \hline
  \end{tabular}
  \tablefoot{Ranges in the radial coordinate used for calculating
    $\chi^2$ during the four consecutive fits. $30\arcsec\approx
    0.03R_\sun$. See also Table~\ref{tab:model-parameters}.} 
\end{table}

We demonstrate the fitting using the 630.2~nm HA scan in the direction
away from DC. The fits to other scans show similar behavior.

\begin{figure}[tbp]
  \centering
  \def\tilewidth{\linewidth}
  \includegraphics[viewport=42 37 710 530,clip,width=\tilewidth]{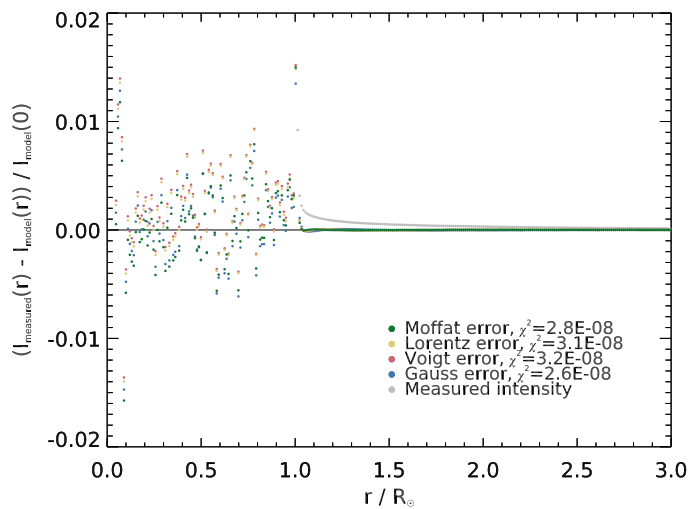}
  \caption{Intensity and fit errors including the disk,
    $N_\text{Gauss}=1$. HA scan at 630~nm, away from DC. The legends
    indicate the kernels used for the widest straylight component.}
  \label{fig:fit_errors_Ngauss}    
\end{figure}

Figure~\ref{fig:fit_errors_Ngauss} shows the fit errors over the
entire radial range. The $\chi^2$ values are dominated by the noise on
the disk. More relevant for our purposes (we are mostly interested in
the off-disk part) are the errors off-disk. Unless stated otherwise,
all $\chi^2$ values mentioned below are calculated for $r>R_\sun +
30\arcsec$.

\begin{figure}[tbp]
  \centering
  \def\tilewidth{\linewidth}
  \subfloat[$N_\text{Gauss}=1$]{\includegraphics[viewport=17 117 695 449,clip,width=\tilewidth]{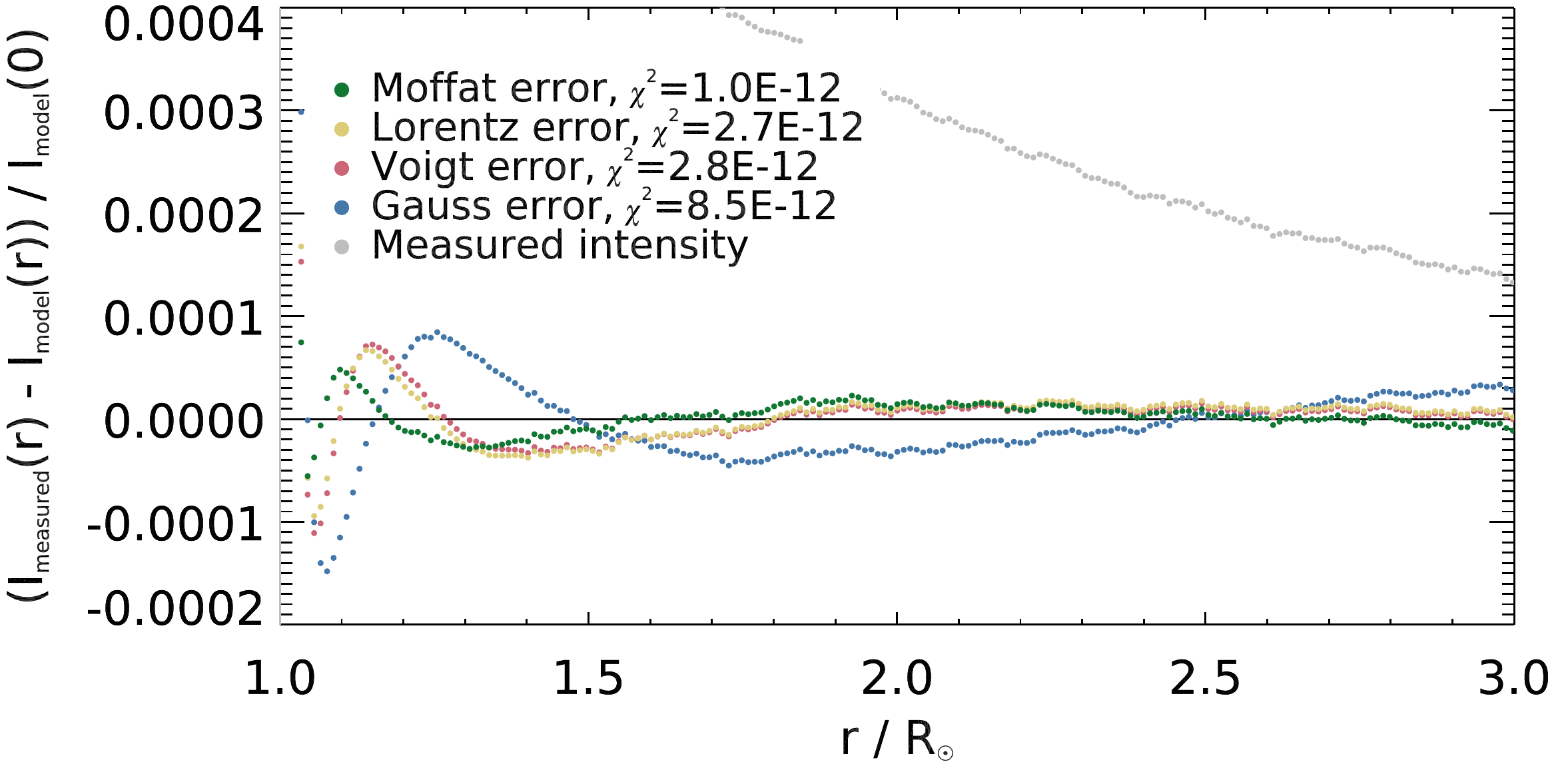}}
  \\
  \subfloat[$N_\text{Gauss}=2$]{\includegraphics[viewport=17 117 695 449,clip,width=\tilewidth]{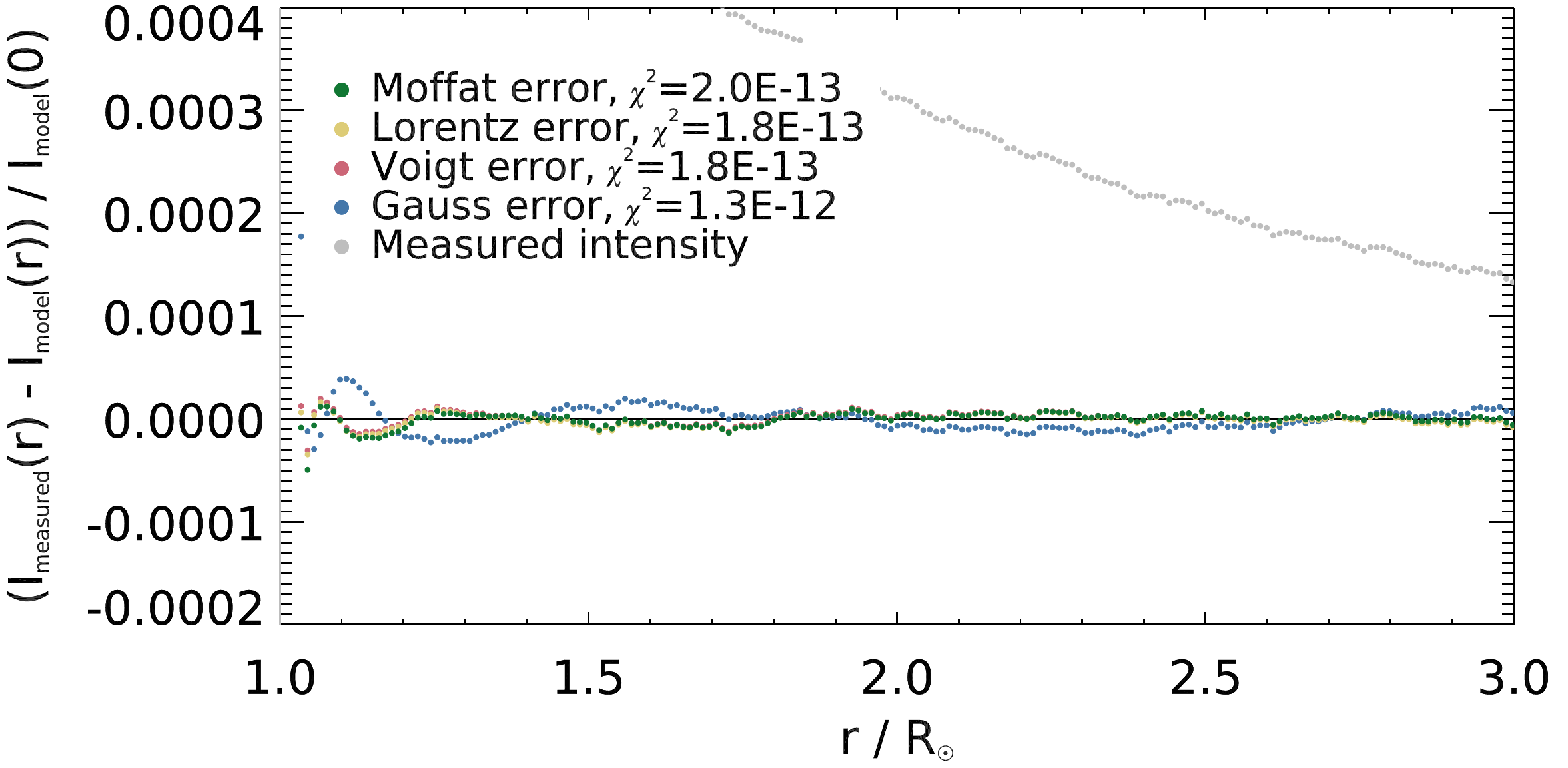}}
  \\
  \subfloat[$N_\text{Gauss}=3$\label{fig:fit_errors_offdisk_Ngauss3}]{\includegraphics[viewport=17 117 695 449,clip,width=\tilewidth]{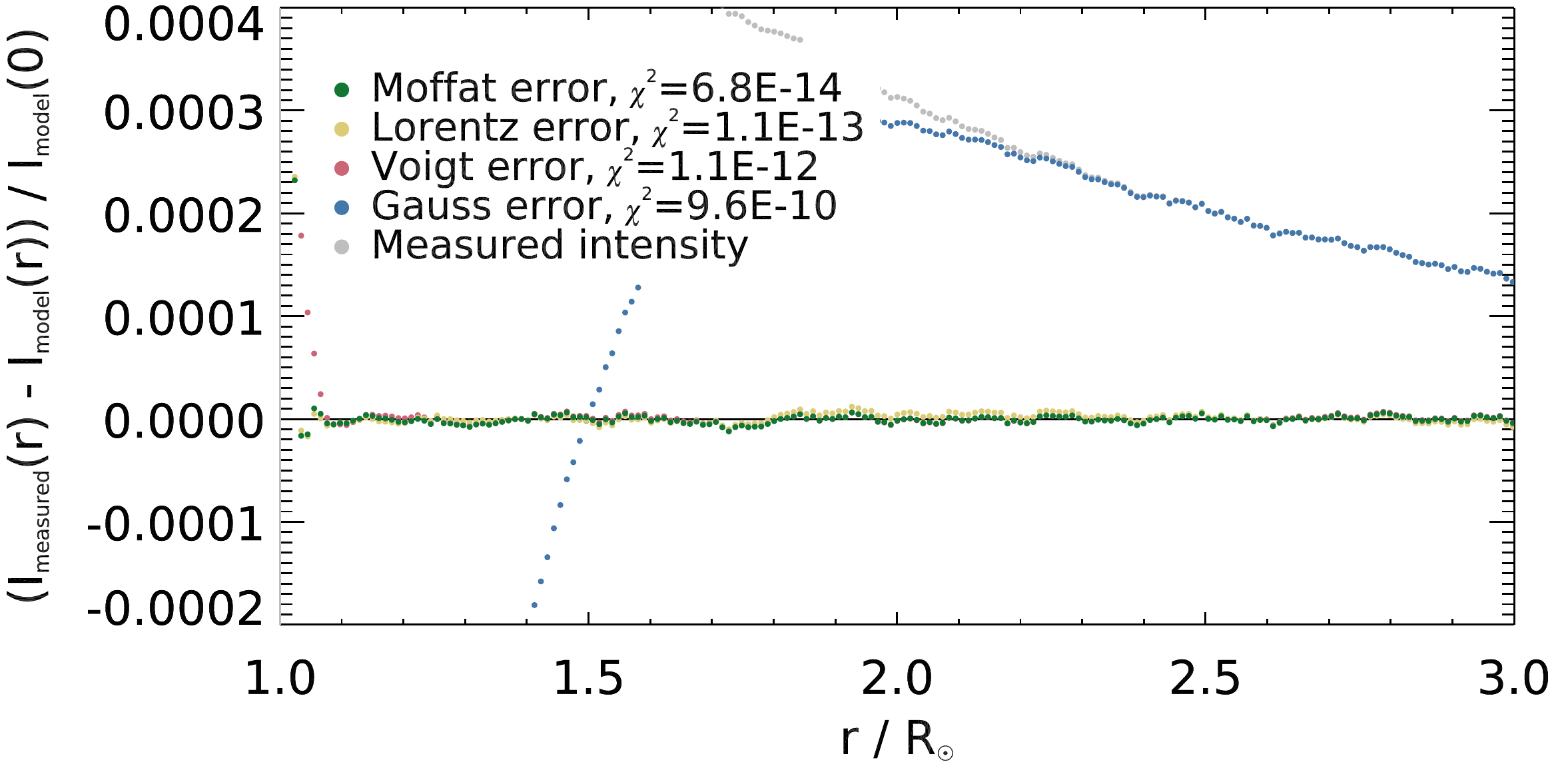}}
  \caption{Fit errors off-disk, different $N_\text{Gauss}$. HA scan at
    630~nm, away from DC. The legends indicate the FWKs used in the
    fits.}
  \label{fig:fit_errors_offdisk_Ngauss}    
\end{figure}

Zooming in on the off-disk errors,
Fig.~\ref{fig:fit_errors_offdisk_Ngauss} demonstrates how the fit
improves with increasing $N_\text{Gauss}$, the number of NGKs.
$N_\text{Gauss}=1$ and $N_\text{Gauss}=2$ show undulating fit errors
but with $N_\text{Gauss}=3$ the errors are dominated by noise for
three of the FWKs (while the Gaussian FWK has failed). This indicates
that there is no point in using more NGKs.

\begin{figure}[!tbp]
  \centering
  \includegraphics[viewport=65 44 700 531,clip,width=\linewidth]{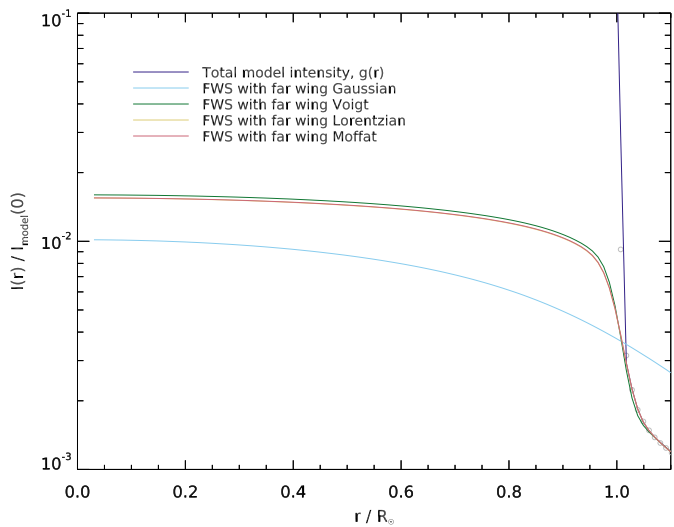}
  \caption{Fitted FWS for 630 nm, $N_\text{Gauss}=3$, HA scan away
    from DC. The total model intensity, $g(r)$, is fitted to the
    measured data represented by the gray circles.}
  \label{fig:straylight-total-allkernels}
\end{figure}

The fit is robust in the sense that, except for the clearly failed
Gaussian FWK, the different FWKs all agree on the total straylight
fraction, as shown in Fig.~\ref{fig:straylight-total-allkernels}.

\begin{figure}[!tbp]
  \centering
  \includegraphics[viewport=65 44 700 531,clip,width=\linewidth]{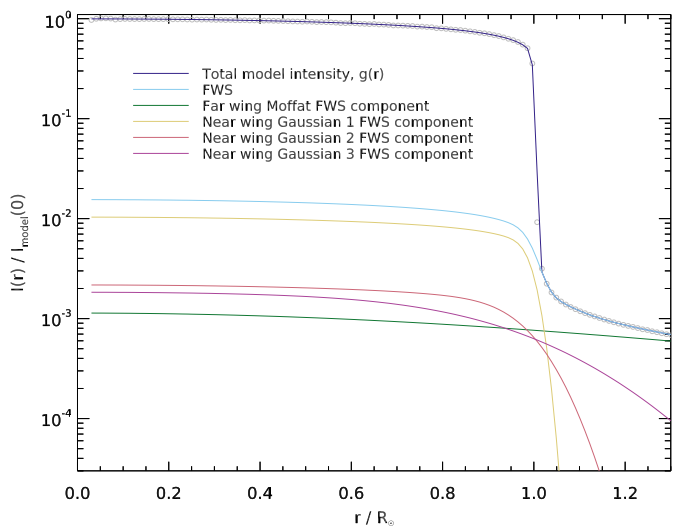}
  \caption{Fitted straylight components for 630 nm,
    $N_\text{Gauss}=3$, Moffat, HA scan away from DC. The total
    model intensity, $g(r)$, is fitted to the measured data
    represented by the gray circles.}
  \label{fig:straylight-components}
\end{figure}

In Fig.~\ref{fig:straylight-components} we show the fitted straylight
decomposed into contributions from the three NGKs and the Moffat FWK.
The dominating component is the most narrow NGKs. Note also the
excellent fit just outside the limb.

\subsection{Results}
\label{sec:results}

The fits result in different combinations of the four kernels (the
three NGKs and one of the four FWKs). Their FWHMs do not stray too
far from their initial values, so the NGK widths are around
50\arcsec{}, between 120\arcsec{} and 160\arcsec{}, and around
400\arcsec{}, respectively. The FWK widths stay within a factor of
about 2 from $R_\sun$. The $c_i$ parameters for the NGKs sum to a few
percent, while $c_0$ for the FWK is usually $\mbox{}\la1$\%.

The width of the blurring kernel, $\textsc{fwhm}_\text{S}\la10\arcsec$
for all wavelengths, which is well within what can be expected from
seeing. It represents the combination of seeing, limb radiative
transfer effects, and all major wavefront straylight with widths of
the order 1\arcsec{}.

We can compare the least wide NGK fits to the unidentified
instrumental scattering from below the telescope field lens found by
\citet{lofdahl12sources}. They measured on the order 0.3\% (0.1\%)
straylight with a width of 20\arcsec{} (34\arcsec) in 395.4~nm
(630.2~nm). Unless the instrumental straylight has increased since
2010 when their data were collected, the now measured straylight must
come from either the telescope itself or the atmosphere.

The sum of $c_i$ is the fraction of energy that is removed from the
direct sunlight. However, the fraction of straylight in observed
images is less because a significant fraction of the straylight is
deposited outside the disk, particularly for the FWKs, which is why we
can measure the straylight there. So the exact values of the fitted
$c_i$ are not as interesting as the total straylight as a fraction of
the direct intensity from the disk. At any point on or off the disk,
the FWS corresponds to the summed contributions to $g(r)$ from
convolution of the limb darkening function, $h(r)$, with the
straylight kernel terms of $k_\text{S}$.

\begin{figure*}[!tbp]
  \centering
  \subfloat[Normalized to unit model intensity
  at DC.\label{fig:straylight-fraction_0}]{\includegraphics[viewport=65 44 700 562,clip,width=0.49\linewidth]{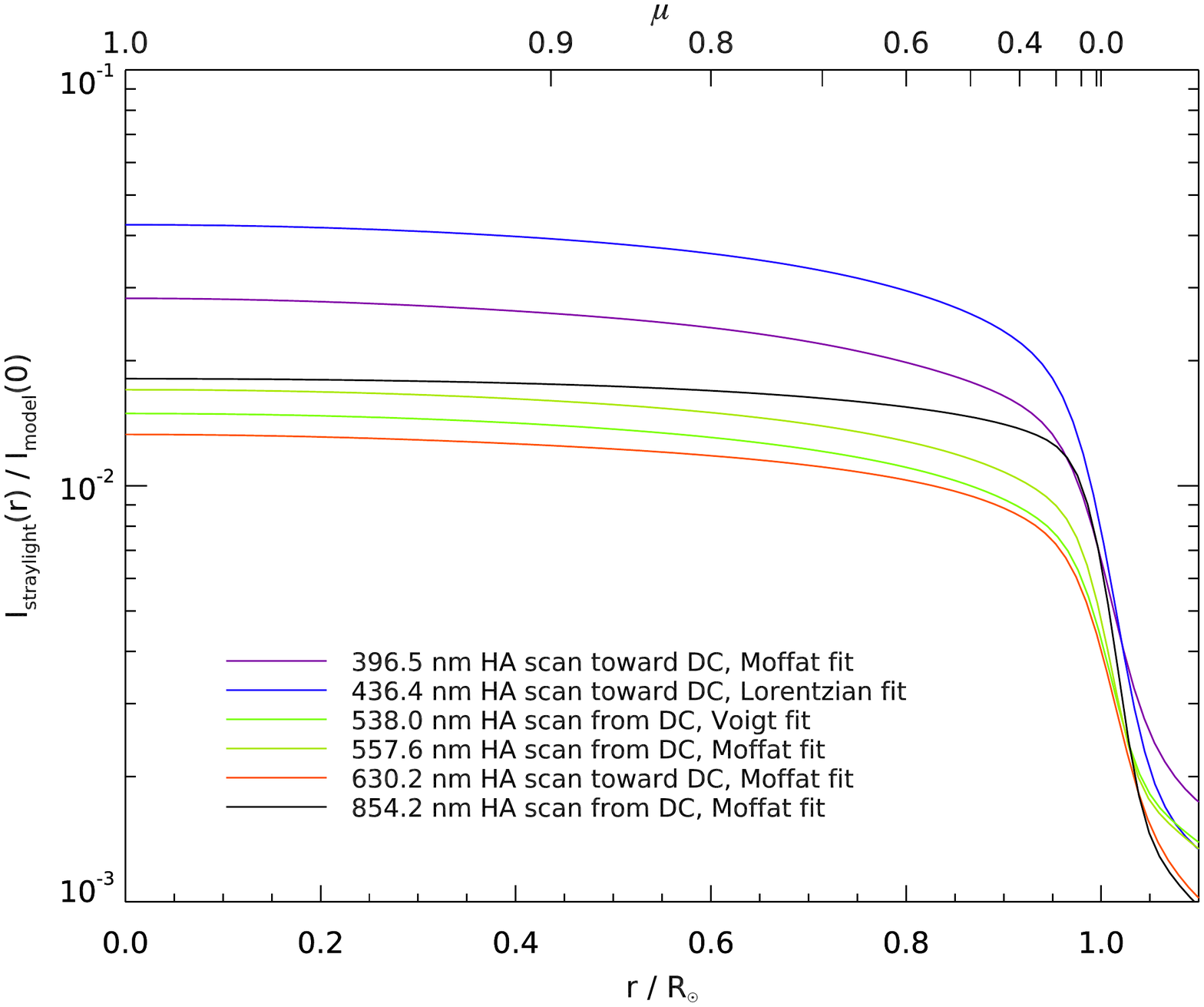}}
\hfill%  \\
  \subfloat[As a fraction of the model
  intensity.\label{fig:straylight-fraction_r}]{\includegraphics[viewport=65 44 700 562,clip,width=0.49\linewidth]{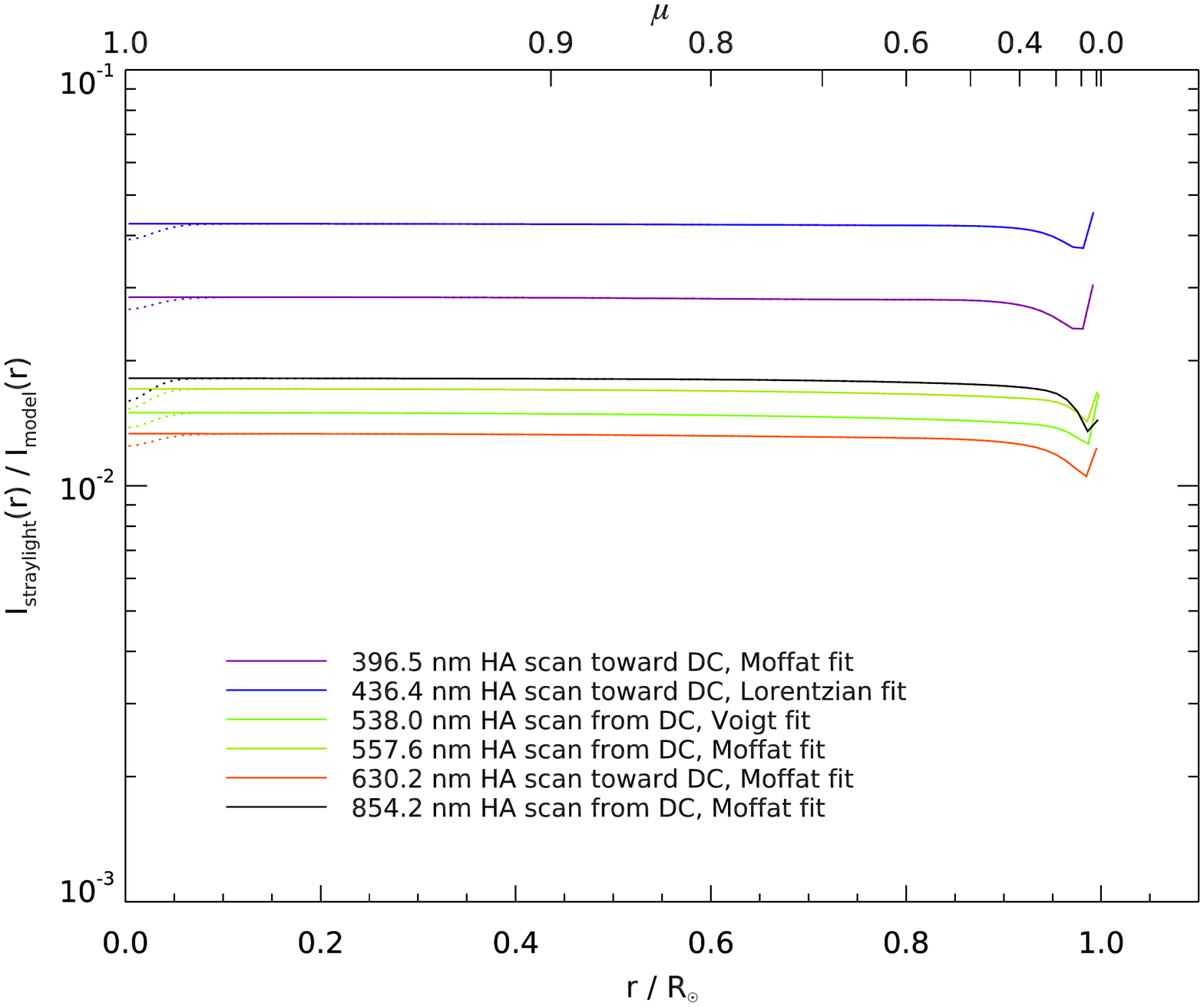}}
  % scatter.pro
  \caption{Fitted FWS for all wavelengths, $N_\text{Gauss}=3$,
    selected HA scan fits. The dotted lines in (b) show the influence
    of an idealized sunspot at DC.}
  \label{fig:straylight-fraction}
\end{figure*}

Figure~\ref{fig:straylight-fraction} shows the total FWS for one
selected HA scan fit per wavelength.
Figure~\ref{fig:straylight-fraction_0} shows it normalized to the DC
intensity, with drops toward the limb due to the limb darkening.

In Fig.~\ref{fig:straylight-fraction_r} the FWS is normalized to the
local intensity, which shows that the straylight is an approximately
constant fraction of the local granulation intensity. The variations
near to the limb are a consequence of the combination of limb
darkening and the fact that the kernels have less disk to collect
energy from there. The dotted lines in this plot show the influence on
the straylight of an idealized sunspot at DC (a zero-intensity umbra
with an arbitrary 13\arcsec{} radius). This changes the straylight by
$\sim$10\% near the spot and not at all at distances $\ga0.05R_\sun$
(or~1\arcmin) from the spot.

\begin{table*}[!tp]
  \centering
  \caption{Total far wing straylight, all scans}
  \label{tab:total-all}
  \begin{tabular}{lllc@{}llc@{}ll}
    \hline\hline\noalign{\smallskip}
    \multirow{2}{*}{$\lambda$} & \multicolumn{2}{c}{HA} && \multicolumn{2}{c}{DA} && \multicolumn{2}{c}{DA extra} \\
    \cline{2-3}    \cline{5-6}    \cline{8-9} \noalign{\smallskip}
    & \multicolumn{1}{c}{toward DC} & \multicolumn{1}{c}{from DC} &&
    \multicolumn{1}{c}{toward DC} & \multicolumn{1}{c}{from DC} &&
    \multicolumn{1}{c}{toward DC} & \multicolumn{1}{c}{from DC} \\
    \hline\noalign{\smallskip}
    396.5 &   2.8 (1.3e--13\,M) &   7.1
    (4.6e--12\,M) &&   2.4 (3.9e--13\,M) &   3.4 (3.3e--13\,V) &&
    2.4 (3.4e--13\,M) &   3.0 (4.1e--13\,V) \\
    436.4 &   4.3 (2.8e--14\,L) &   5.2
    (2.1e--14\,M) &&   2.9 (3.4e--13\,M) &   6.5 (6.3e--12\,V) &&
    3.4 (2.4e--13\,M) &   8.4 (5.4e--12\,V)  \\
    538.0 &   1.5 (6.3e--13\,V) &   1.5
    (9.1e--14\,V) &&&&&& \\
    557.6 &   1.8 (5.5e--13\,M) &  1.7
    (8.3e--14\,M) &&   0.7 (7.1e--13\,M) &   1.3 (3.9e--13\,M) &&& \\
    630.2 &  1.3 (3.5e--13\,M) &   1.5
    (6.8e--14\,M) &&   0.7 (4.4e--13\,M) &  \llap{5}0.4 (4.1e--07\,M)
    &&& \\
    854.2 &   1.3 (9.4e--13\,L) &   1.8
    (1.7e--13\,M) &&   1.1 (9.8e--13\,L) &   0.7 (8.4e--13\,G) &&
    1.0 (1.8e--12\,M) &   1.6 (1.0e--12\,V) \\
    \hline
  \end{tabular}
  \tablefoot{Each cell shows the best estimate straylight fraction in
    percent and in parentheses the outside-the-disk $\chi^2$ of the
    best fit and the initial of the kernel that gave the best fit. The
    reported straylight fraction is the median on-disk fraction,
    compare Fig.~\ref{fig:straylight-fraction_r}.  The DA ``extra''
    data are from the 11:49 scans.} 
\end{table*}

Table~\ref{tab:total-all} summarizes the results of fitting the model
to all our data sets, including the 11:49 ``extra'' scans. Comparing
with Fig.~\ref{fig:wings}, we can note the following (with $\pm$
simply denoting the extremes of the estimates):
\begin{description}
\item[396.5~nm:] These highly consistent scans correspond to consistent
  estimates of straylight, $2.9\pm0.5$\%. The exception is the HA FDH,
  for which the estimate is more than twice as high but $\chi^2$ is
  also higher because of the upward glitch at $r=2.4R_\sun$.
\item[436.4~nm:] These scans also look consistent, except for the two
  DA TDHs that both end with downward glitches at $r\approx2.3R_\sun$.
  Disregarding the lower estimates of the exceptions, we get
  $6.3\pm2$\%. 
\item[538.0~nm:] These two scan halves look similar and the estimates
  are the same, 1.5\%. The inconsistencies at $r<1.4R_\sun$ do not
  appear to influence the estimate much. The leveling off at
  $r>3R_\sun$ does not seem to have been a problem.
\item[557.6~nm:] High consistency between the two halves of the HA
  scan result in the estimates $1.75\pm0.05$\%. The DA scan is noisier
  and lower than the HA scan and the halves end at different $r$,
  resulting in the estimate $1.0\pm0.3$\%.
\item[630.2~nm:] These scans look similar to the 557.6~nm scans. The
  estimated stray light levels are lower than for 557.6~nm due to the
  LD not falling off as quickly near the limb. The HA scan estimate is
  consequently $1.4\pm0.1$\%. Consistently, the DA TDH estimate is
  0.7\%. The DA FDH results can be disregarded, as the fits failed
  completely for all FWKs with the best $\chi^2$ several orders of
  magnitudes larger than for any other data.
\item[854.2~nm:] The filter with the most inconsistent scans and scan
  halves, and also the noisiest DA scans. The HA estimate is
  consistent with the levels in the plot: $1.55\pm0.25$\% with the FDH
  higher than the TDH. The DA scan results are less consistent with
  $0.9\pm0.2$\% for DA and $1.3\pm0.3$\% for DA ``extra''.
\end{description}

\subsection{The PSFs}
\label{sec:psfs}

\begin{figure*}[p]
  \centering
  \def\tilewidth{0.495\linewidth}
  \subfloat[396.5 nm]{\includegraphics[viewport=50 44 700 528,clip,width=\tilewidth]{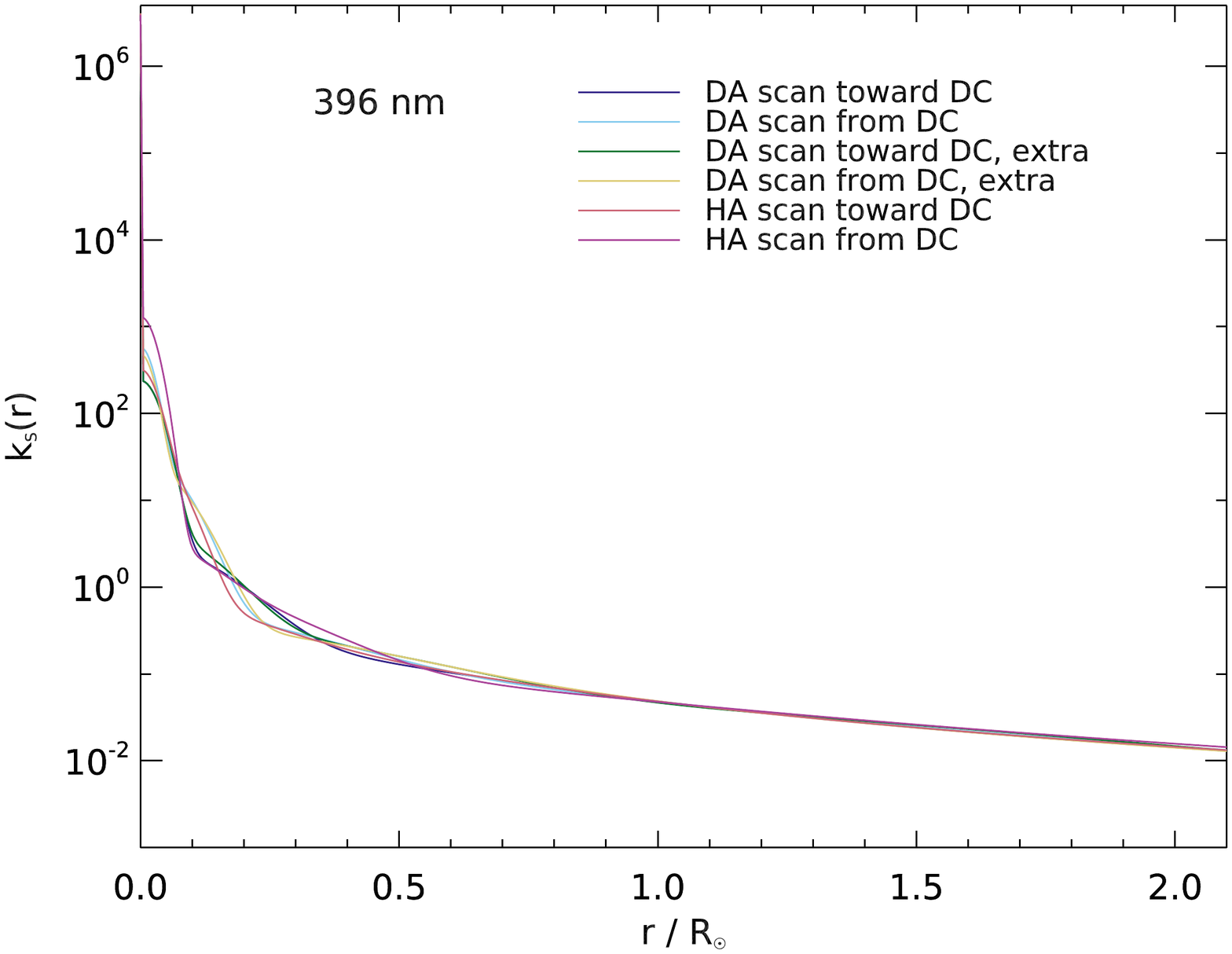}}\hfill
  \subfloat[436.4 nm]{\includegraphics[viewport=50 44 700 528,clip,width=\tilewidth]{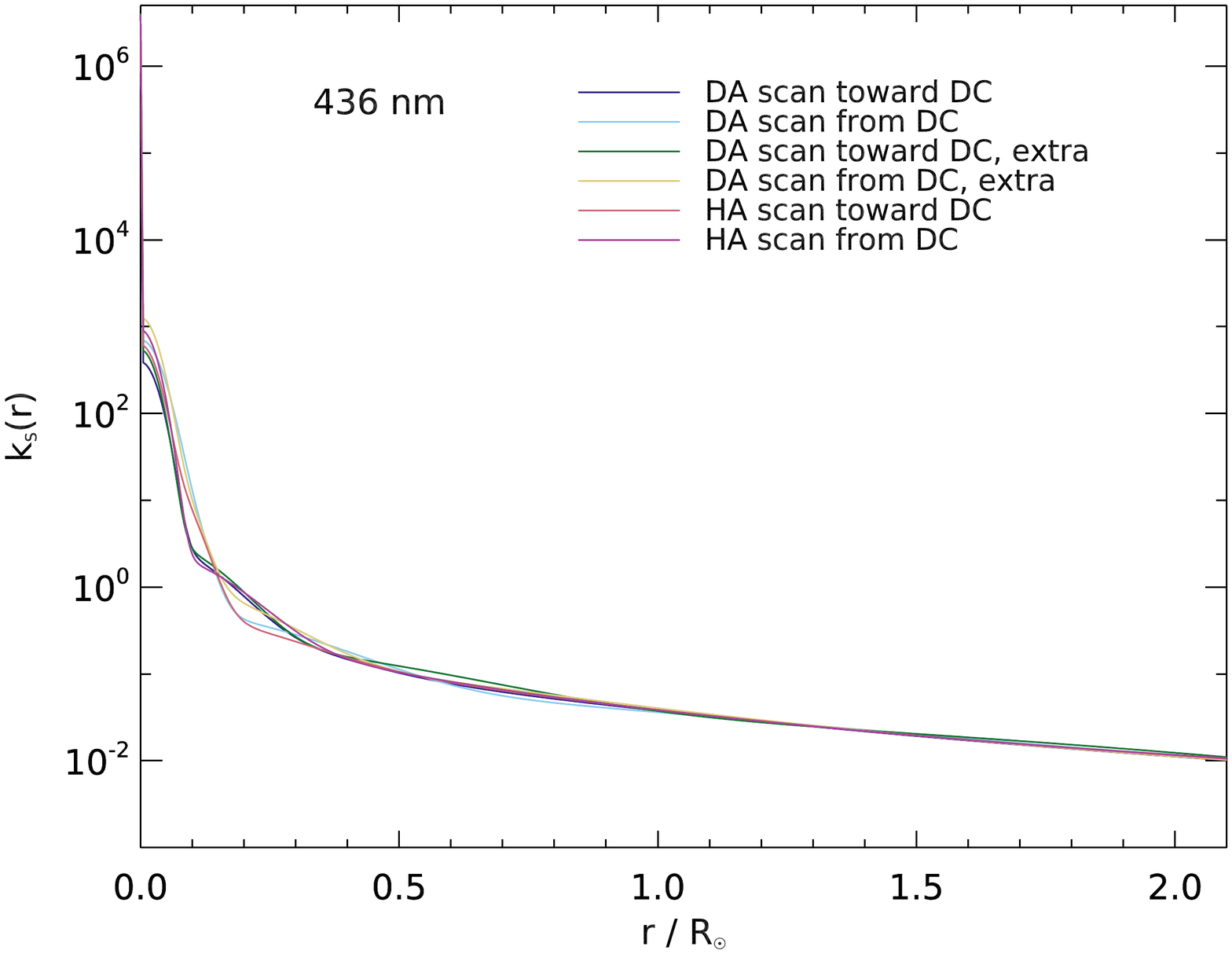}}\\
  \subfloat[538.0 nm]{\includegraphics[viewport=50 44 700 528,clip,width=\tilewidth]{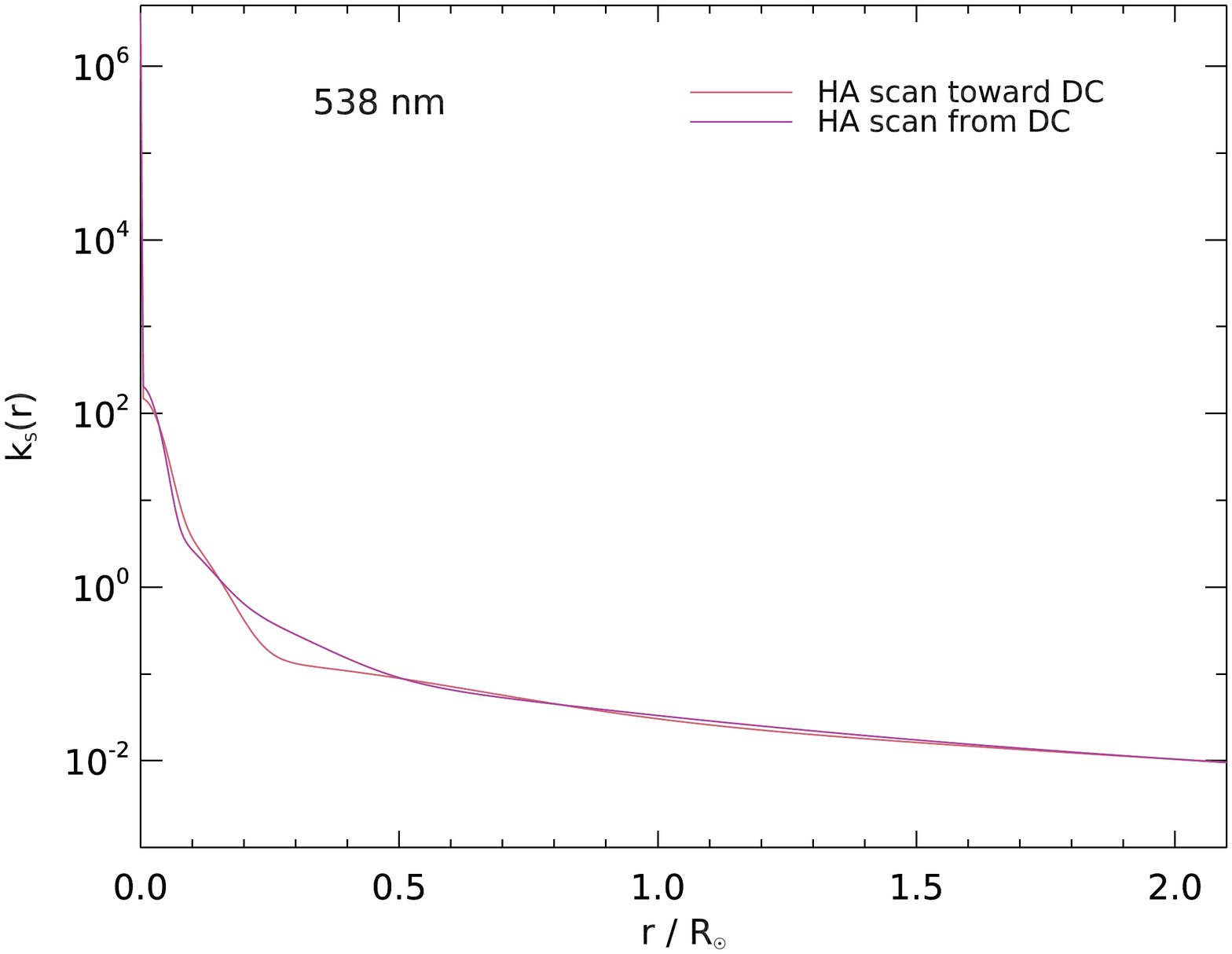}}\hfill
  \subfloat[557.6 nm]{\includegraphics[viewport=50 44 700 528,clip,width=\tilewidth]{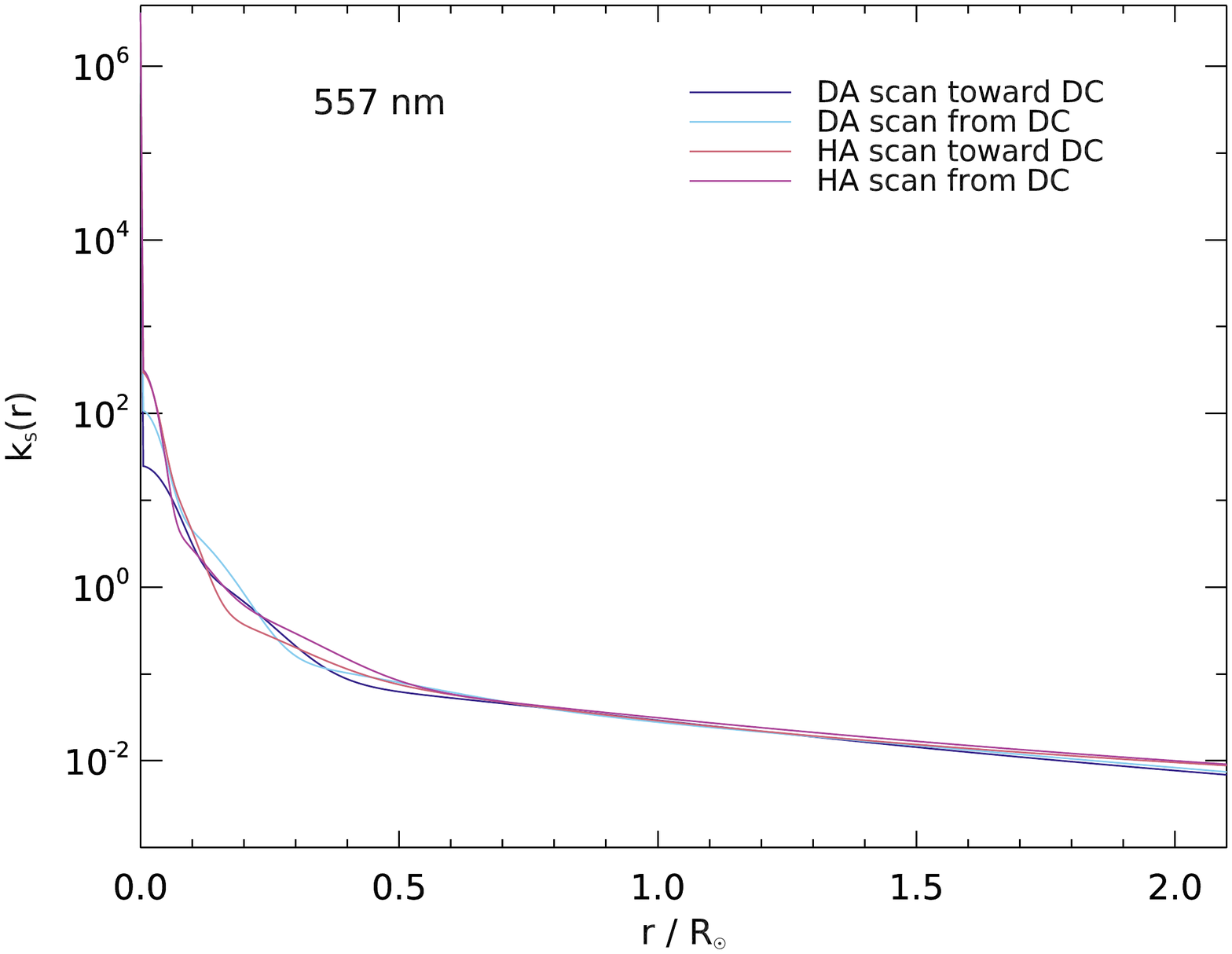}}\\
  \subfloat[630.2 nm]{\includegraphics[viewport=50 44 700 528,clip,width=\tilewidth]{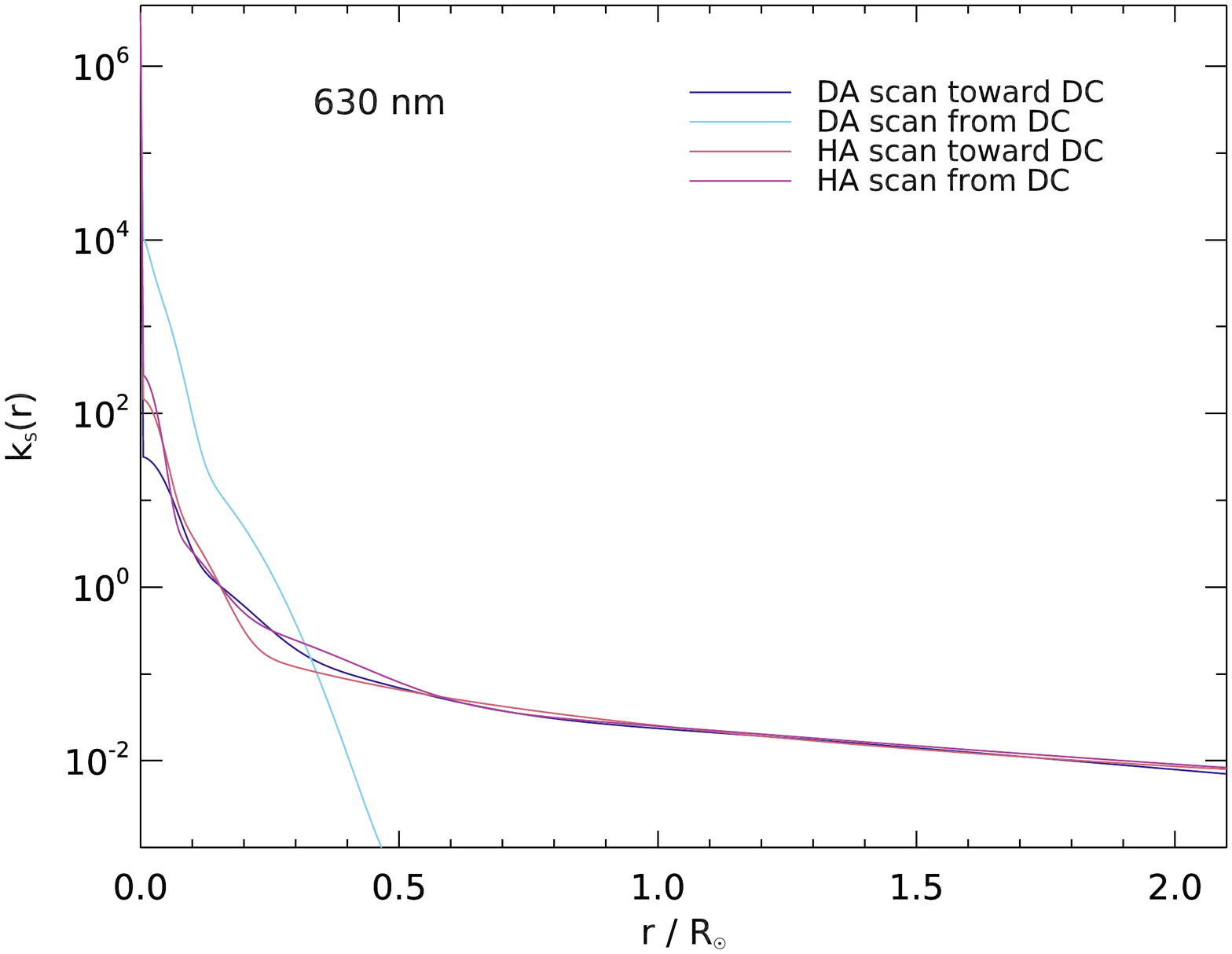}}\hfill
  \subfloat[854.2 nm]{\includegraphics[viewport=50 44 700 528,clip,width=\tilewidth]{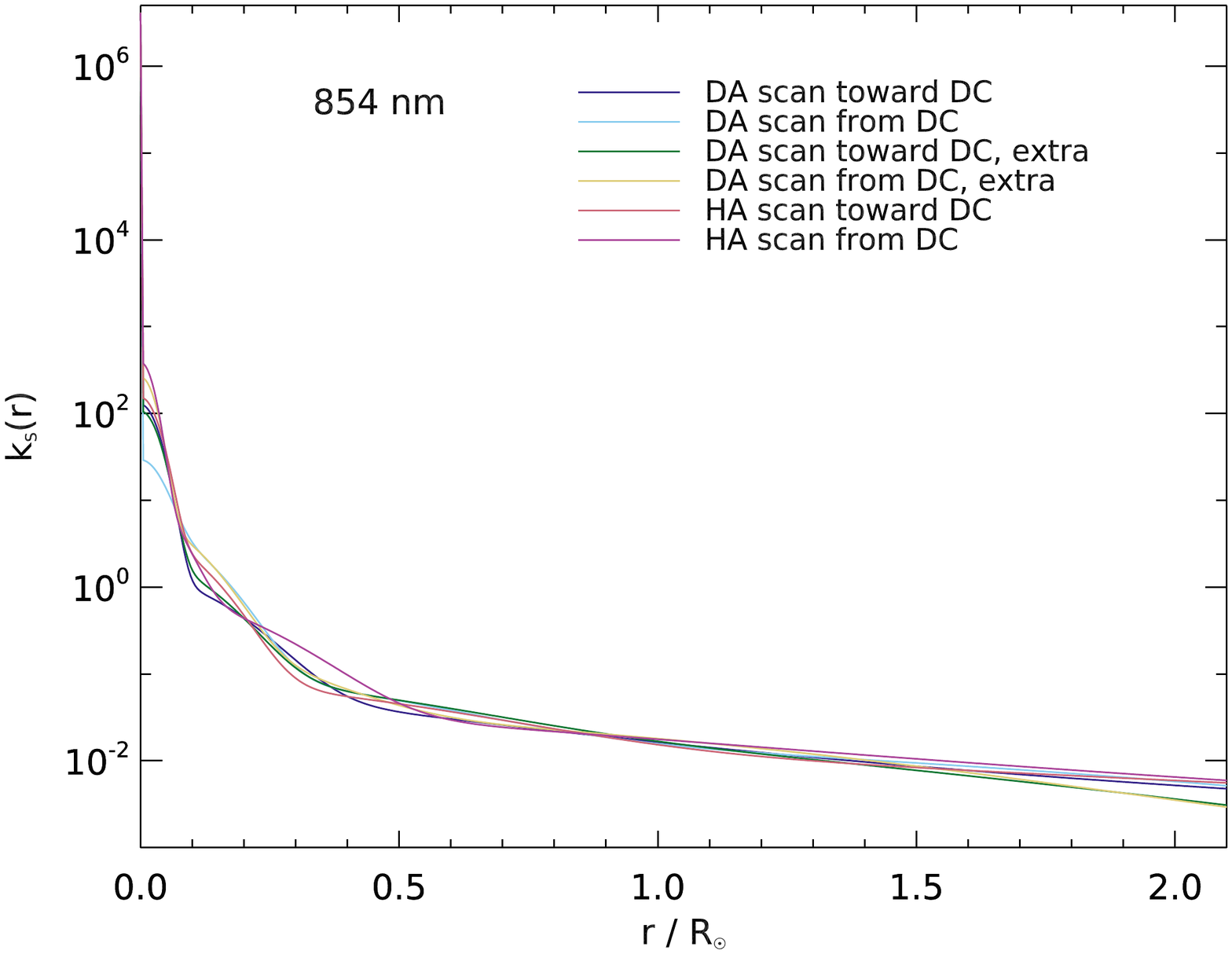}}\\
  \caption{Scattering kernels $k_\text{S}$. Compare
    Fig.~\ref{fig:wings} and Table~\ref{tab:total-all}.}
  \label{fig:psfs}
\end{figure*}

Figure~\ref{fig:psfs} shows the fitted PSFs. While the straylight at
large $r$ differs from scan to scan in each of the plots in
Fig.~\ref{fig:wings}, the far wings of the fitted scattering kernels
are remarkably consistent. The variation is mostly in the
$r\la0.5R_\sun$ range.

\begin{figure*}[p]
  \centering
  \def\tilewidth{0.495\linewidth}
  \subfloat[396.5 nm]{\includegraphics[viewport=50 44 710 528,clip,width=\tilewidth]{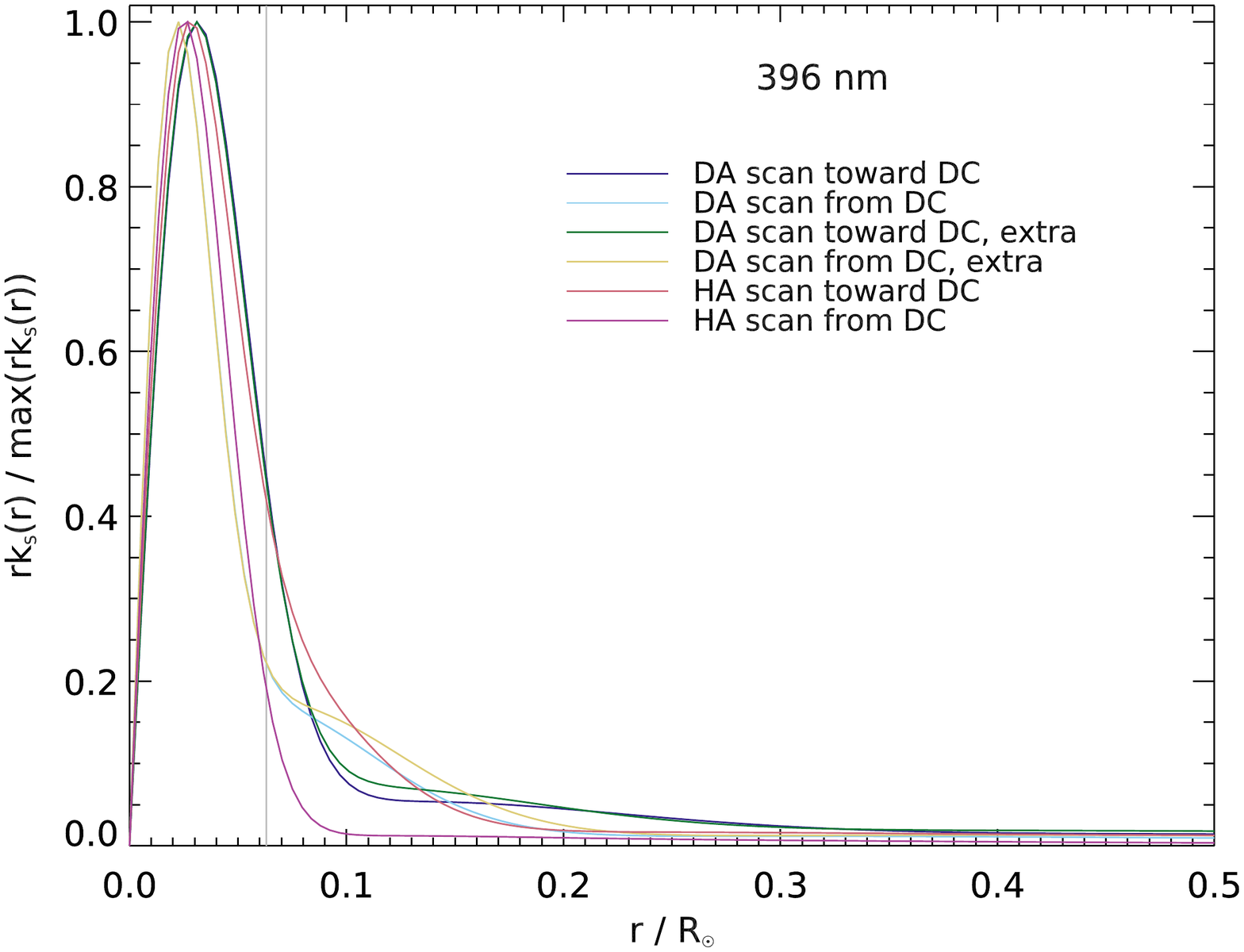}}\hfill
  \subfloat[436.4 nm]{\includegraphics[viewport=50 44 710 528,clip,width=\tilewidth]{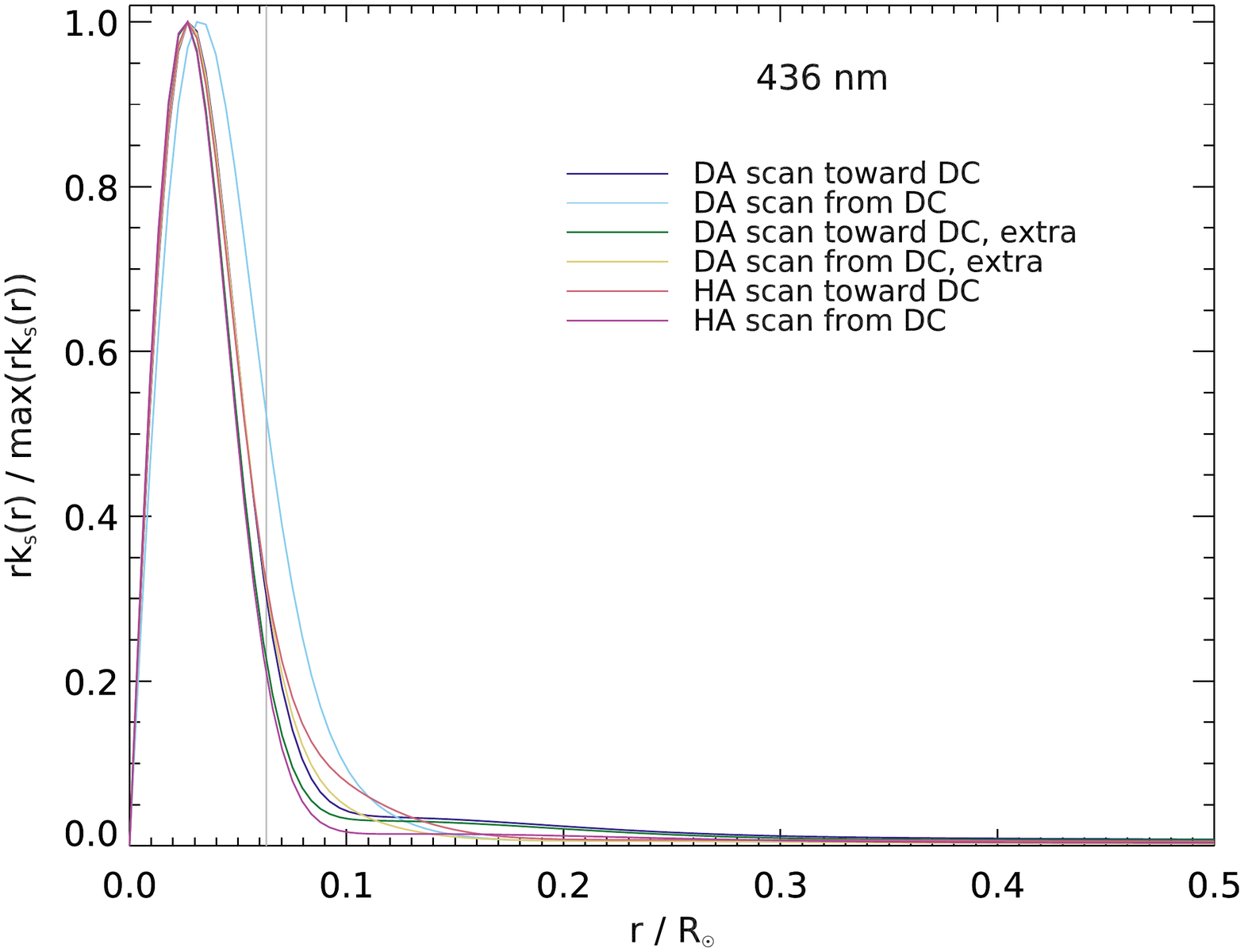}}\\
  \subfloat[538.0 nm]{\includegraphics[viewport=50 44 710 528,clip,width=\tilewidth]{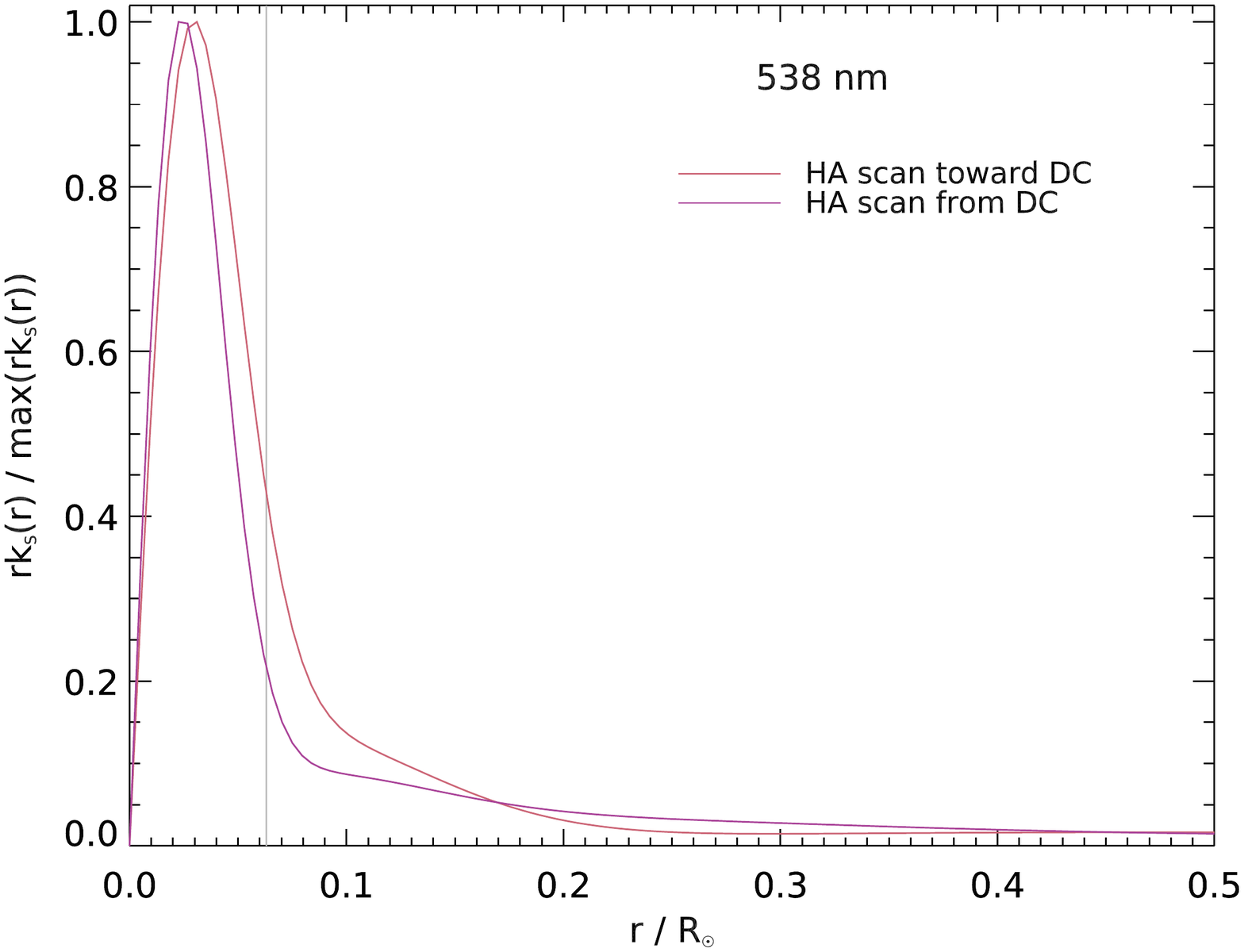}}\hfill
  \subfloat[557.6 nm]{\includegraphics[viewport=50 44 710 528,clip,width=\tilewidth]{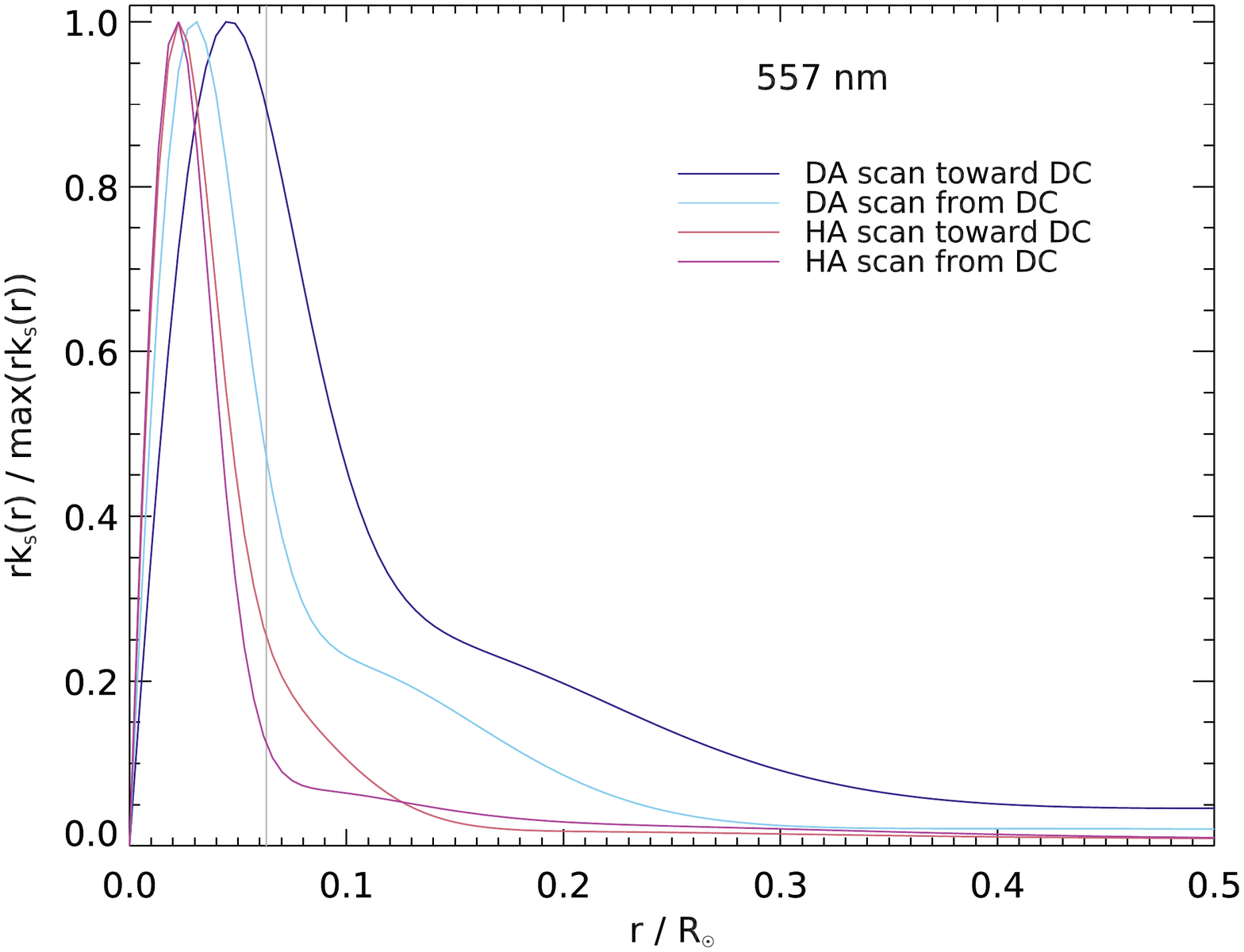}}\\
  \subfloat[630.2 nm]{\includegraphics[viewport=50 44 710 528,clip,width=\tilewidth]{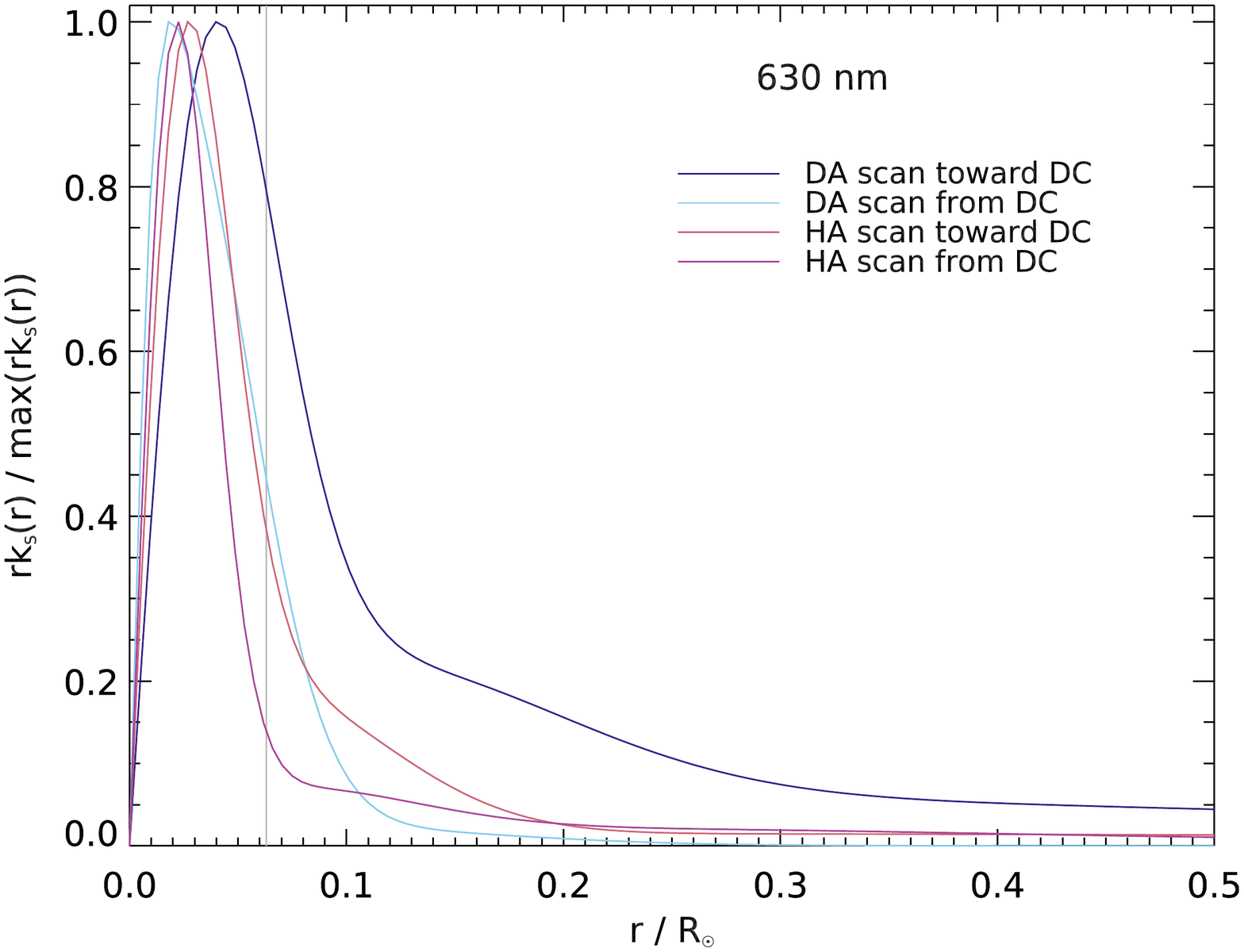}}\hfill
  \subfloat[854.2 nm]{\includegraphics[viewport=50 44 710 528,clip,width=\tilewidth]{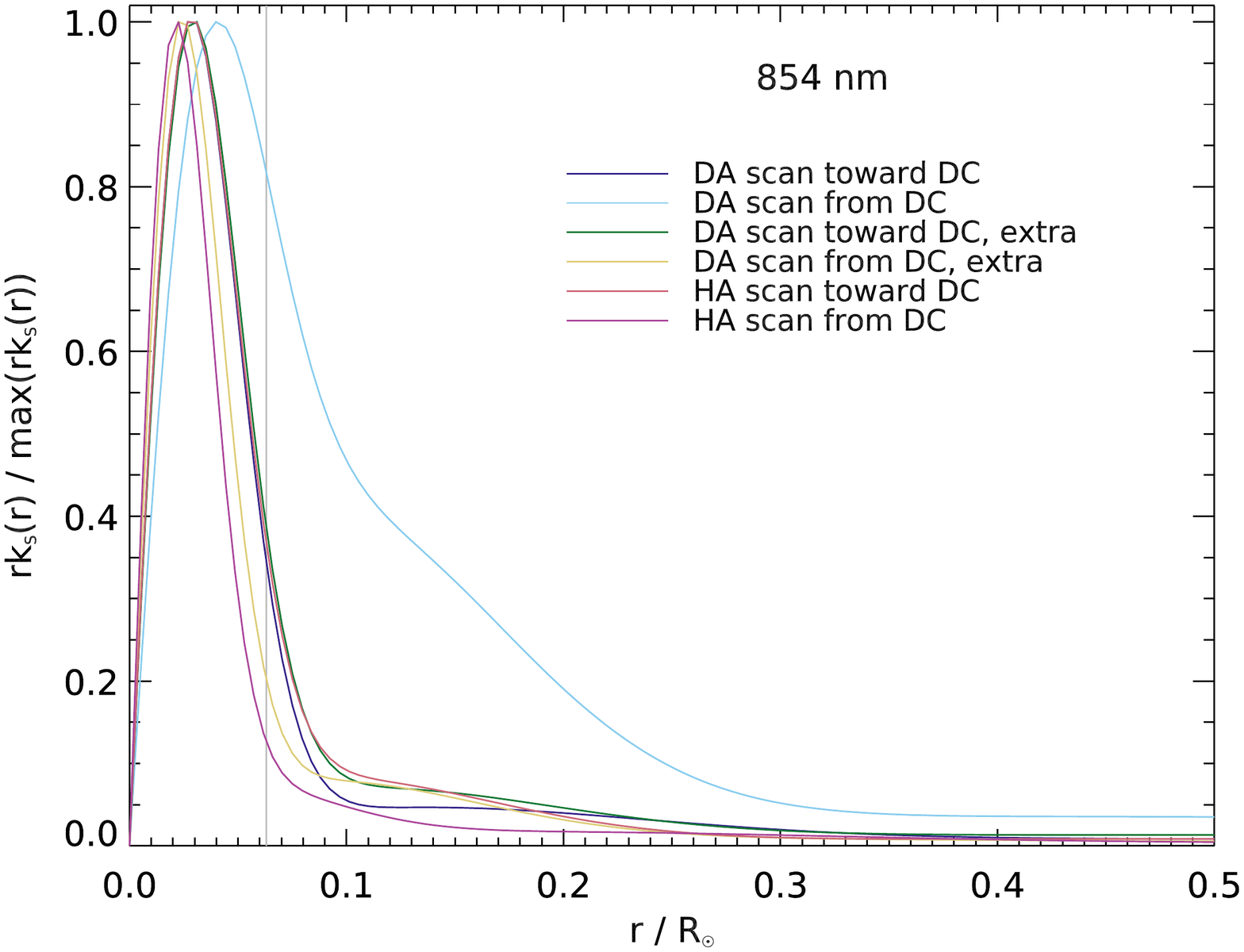}}\\
  \caption{Scattering kernels radial contributions, $rk_\text{S}$.
    Individually normalized to unit maximum. The vertical lines
    indicate $r=1\arcmin$. Compare Fig.~\ref{fig:psfs}.}
  \label{fig:psfs_r}
\end{figure*}

In Fig.~\ref{fig:psfs_r} we plot the PSFs times the radial coordinate,
i.e., weighted by the annular surface area from which energy is
redistributed by convolution. These plots show that the bulk of the FWS
is scattered from within $r\la 1\arcmin$.

\section{Conclusion} 
\label{sec:conclusion}

We have measured the aureola intensity of the SST, out to $3R_\sun$
and in several wavelengths. This allowed us to characterize the far
wing straylight (FWS) PSFs, including the wings far enough to capture
all stray light that can influence science image data on the entire
disk. The measurements were done on a clear day without calima, with
low amounts of dust in the atmosphere (TNG dust count: $\sim$$10^{-1}\
\upmu$g\,m$^{-3}$).

The FWS PSFs scatter light primarily from within $r\la 1\arcmin$ but
have tails that extend to several solar radii.

Locally, the FWS of the SST is an approximately constant addition to
science images of a few percent of the surrounding granulation
intensity, much less than needed to explain the discrepancy between
granulation contrast in observed and synthetic images. Particularly in
CRISP data, the FWS is only $\la 2$\%. Correcting RMS contrast
measurements for this would increase the contrast by the same
fraction, such as from 10.0\% of the average intensity to 10.2\%. This
is insignificant compared to the deficiencies in contrast of
granulation data that are compensated for seeing effects by AO, MFBD
restoration, and post-restoration deconvolution for uncompensated
high-order wavefront modes reported by \cite{scharmer10high-order}.
They measured 10\% and 8.5\% RMS contrast in 538~nm and 630~nm,
respectively, while MHD synthetic data at the same viewing angle has
17.4\% and 13.3\% RMS contrast in the same wavelengths.

The measured FWS varies with wavelength and time, most likely due to a
combination of varying zenith angle and variations in the dust
concentration. Using scans from several days and zenith angles,
\citet{1990SoPh..125..211M} were able to separate the atmospheric
scattering from the instrumental scattering of the Vacuum Newton
Telescope and showed that the instrumental scattering was
approximately constant (their Fig.~3). We do not have enough data for
a similar separation. However, the very minor variations with zenith
angle in the blue data are consistent with the measured FWS being
dominated by non-varying scattering in the telescope. Allowing for an
atmospheric contribution (certainly present for the redder CRISP
wavelengths), the measured FWS represents an upper limit for the
instrumental scattering of the SST.

Acquisitions of this kind of calibration data can be made in
$\la10$~min per CRISP prefilter (see Table\ref{tab:observations}), so
FWS calibrations could be made routinely. As currently implemented,
the calculations are quite time consuming but could be stream-lined
and ported to C for speed and added as a voluntary step in the
CRISPRED data reduction pipeline of \citet{delacruz15crispred}.
However, as SST data are mostly collected during low-dust conditions
(calima often comes with hot weather and bad seeing) such calibrations
would be routinely useful only when the more dominating sources of
contrast-reducing straylight have been fully characterized.

The fact that the off-disk straylight is clearly above the noise level
even in low dust conditions means the calibration should be even
easier to measure in dustier conditions. Still, we did have problems
making the models fit. If similar measurements are to be repeated, one
may consider taking more data, i.e., scan slower than the Sun moves
during a HA drift scan. This would be particularly useful near the
limbs, where the intensity gradient is large. Collecting a limb image
from at least one more position along the limb would make the
coordinate transformations of Section~\ref{sec:coord-transf} more well
determined.

\balance 

\begin{acknowledgements}
  The Swedish 1-m Solar Telescope is operated on the island of La
  Palma by the Institute for Solar Physics of Stockholm University in
  the Spanish Observatorio del Roque de los Muchachos of the Instituto
  de Astrof{\'\i}sica de Canarias. The author is grateful to Mikael
  Ingemyr and Pit S\"utterlin for help with the observations, to Guus
  Sliepen for making it possible to do the DA scans, and to G\"oran
  Scharmer for discussions and helpful suggestions. 
\end{acknowledgements}

%\bibliography{bib-strings_aa,ads,svst,mats,mats.lofdahl}

\end{document}